\documentclass[final,5p,times,twocolumn]{elsarticle} %1p / 5p
\DeclareMathAlphabet{\mathpzc}{OT1}{pzc}{m}{it}
\usepackage{lineno}
\usepackage{setspace}
\usepackage{graphicx}
\usepackage{amssymb}
\usepackage{amsmath}
\usepackage{amsthm}
\usepackage[colorlinks=true]{hyperref}
\usepackage[all]{hypcap}
\usepackage{mathrsfs}

\usepackage[percent]{overpic}
\usepackage{fixltx2e}
\usepackage{epsfig}

\journal{Nuclear Instruments and Methods A}
\begin{document}
\begin{frontmatter}
\title{A coincidental timing model for the scintillating fibers}
\author{Petar \v{Z}ugec\corref{cor1}}
\ead{pzugec@phy.hr}
\cortext[cor1]{Corresponding author. Tel.: +385 1 4605552}
\address{Department of Physics, Faculty of Science, University of Zagreb, Bijeni\v{c}ka cesta 32, Zagreb, Croatia}

\begin{abstract}
A model describing the coincidental timing of scintillating fibers is developed. Fiber geometry, the rate of scintillation decay together with the mean number, spatial dispersion and attenuation of emitted photons is considered. For a specific selection of probability distributions and parameters involved, the entire coincidental timing distributions, corresponding FWHM values and the photon detection efficiencies are extracted. The significance of the number of photons from the scintillation process is specially emphasized. Additionally, the model is extended to include a triggering feature, experimentally realized by coupling fibers to any photon resolving device. Finally, the measurements of a coincidental timing distribution were performed, with an excellent agreement found between the experimental and predicted theoretical results.
\end{abstract}

\end{frontmatter}

\section{Introduction}
\label{sec:chap1}
Scintillating fibers have been extensively investigated from both theoretical and experimental point of view. Consequently, their scintillation mechanisms and light transmission properties are well known and conveniently summarized in \cite{fib1}, together with a list of the most relevant references on the subject. Nowadays, there is an active effort to include fibers as an integral part of sophisticated detector systems for nuclear experiments. Being intended for a detection of charged particles and/or electromagnetic radiation, fibers are required to provide an adequate, if not excellent coincidental timing resolution, vital for precise reconstruction of the scintillating pulse position along their length. Compared to the resolution of bulk scintillators -- an order of magnitude below the nanosecond scale \cite{fib2} -- early measurements indicate that attainable values are well above this range \cite{fib3}. The rate of scintillation process, together with a low number of detected photons being subject to a wide spatial dispersion, is considered to be the main cause for the coincidental timing discrepancy with that of bulk scintillators. Commonly, Monte Carlo simulations present a natural tool for estimation and prediction of experimental results, easily capable of including all the required physical principles. However, in this paper an analytical model is developed, describing an asymptotic form of otherwise measured coincidental timing distributions, giving rise to the central resolution defining FWHM value. Compared to the earlier, long-established models \cite{fiba, fibb}, the one developed herein extends beyond the sole scintillation process, taking into account a subsequent light propagation, while proposing the simple manner in which to include even the effects of the photon detecting units. An assumption of a fully general form for the emission of the scintillation photons, their spatial distribution and subsequent attenuation inside the fiber material allows physical considerations of varying complexity to be adopted -- from a simple meridional approximation to a considerably more refined description of optical processes involved. Due to the level of generality acchieved, though developed with scintillating fibers in mind, the model is indiscriminately applicable to any kind of light guides, making it an acute mathematical tool for predicting the experimental results far beyond the assumptive limitations of a specific setup.  As a starting point a model for the photon propagation times is considered -- initially developed in \cite{fib2} for bulk scintillators and later applied to scintillating fibers in \cite{fib4} -- from which the technical formalism was adopted. 

\section{Basic model}
\label{sec:chap2}
In common nuclear or particle physics experiment the arrival time of a signal is determined by the leading edge or the constant fraction discrimination of the signal's leading edge. Since the first photon impinging on the detector initiates the rise of the signal, its statistics is of the utmost importance for the description of signal timing properties.

\begin{figure}[t] 
\centering 
\includegraphics[width=0.5\textwidth,keepaspectratio]{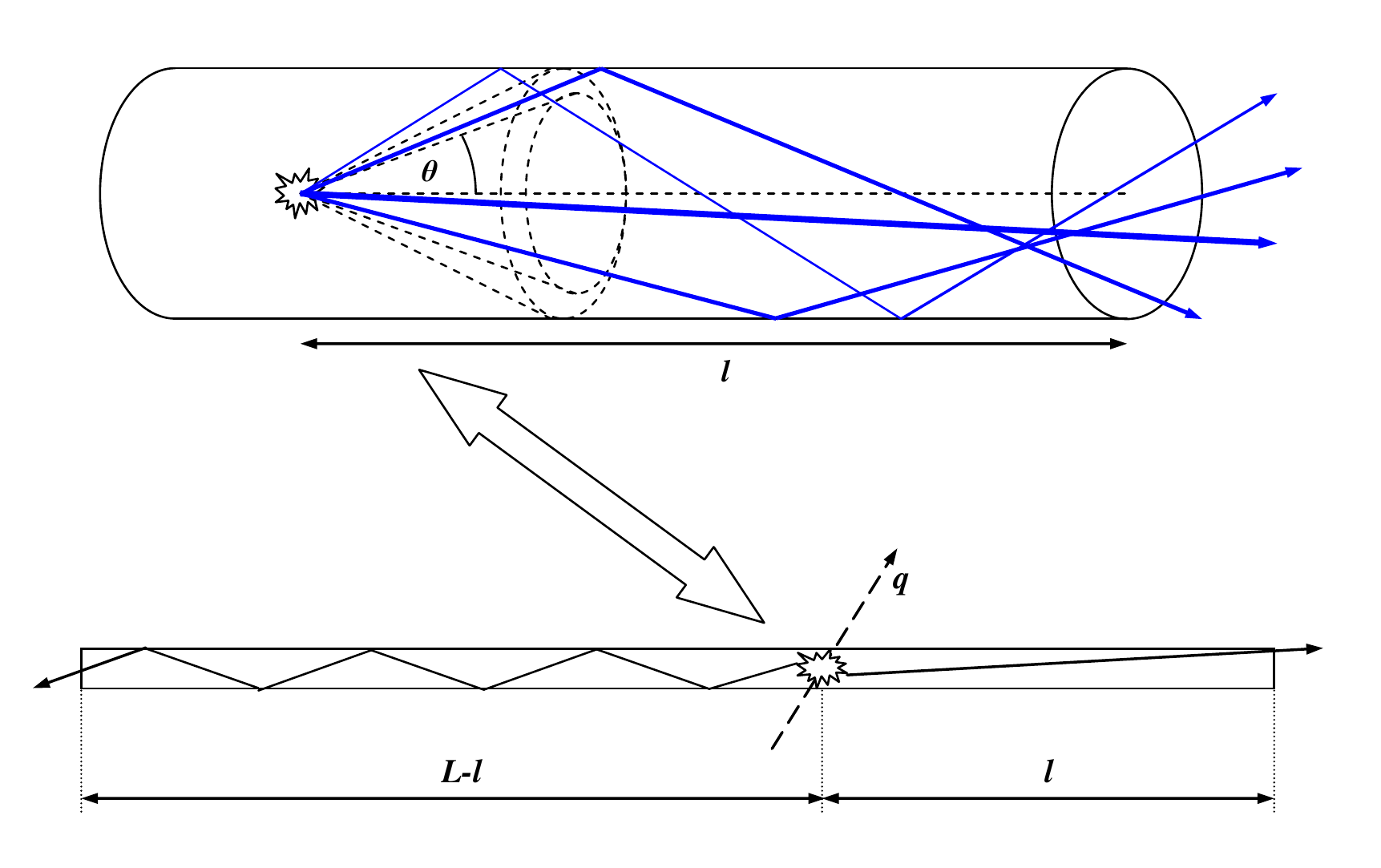}
\caption{A scintillating pulse induced by the charged particle $q$ passing through the fiber at a distance $l$ from one of its ends. The difference of initial photons' times of arrival at each end constitutes a basis for measuring the coincidental timing resolution. A wide spatial dispersion of isotropically emitted photons is also illustrated, increasing the emission probability for greater $\theta$ angles, while degrading attainable resolution.}
\label{fig:fig1}
\end{figure}

Therefore, let us assume that inside the scintillating fiber of a length $L$ at a distance $l$ from one of its ends a scintillation pulse was induced (\autoref{fig:fig1}), emitting a total of $n$ photons. To determine a probability $f_n$ for the first arriving photon to reach the fiber end, several separate cases must be considered. For example, first arriving photon may correspond to the first emitted (probability $p_1$). On the other hand, first $m-1$ emitted ones may be lost either by escaping from the fiber before reaching its end or by absorption inside the fiber material. Therefore, $m$-th emitted one becomes first to reach the fiber end (probability $p_m$) and induce the signal in the detector. Consequently, all such contributions give rise to probability $f_n$:
\begin{linenomath*}\begin{equation*}
f_n=\sum_{m=1}^n p_m
\tag{2.1}
\end{equation*}\end{linenomath*}

Let us beforehand define the moment $t_A$ for the first photon arrival at the fiber end, the moment $t_E$ of its emission and the time $t_P$ needed for its propagation through the fiber. The following holds:
\begin{linenomath*}\begin{equation*}
t_A=t_E+t_P
\tag{2.2}
\end{equation*}\end{linenomath*}
Most simplistic in form, previous relation will be fundamental for final calculations. For a detailed analysis, let us separate contributions to $p_m$:
\begin{linenomath*}\begin{equation*}
p_m=\alpha_m\beta_m\gamma_m
\tag{2.3}
\end{equation*}\end{linenomath*}
with $\alpha_m$ regulating $m-1$ photons emitted before the $m$-th one, $\beta_m$ regulating $m$-th emitted one as first arriving to the fiber end, and $\gamma_m$ regulating all the $n-m$ following ones. For a description of these probability contributions let us assume a completely general form $E(t_E)$ for the emission probability distribution and $S(t_P)$ for the spatial distribution of emitted photons expressed in terms of the photon time propagation, i.e. translated into the path length dispersion. Furthermore, an attenuation factor $A(t_P)$ governing the probability of single photon reaching the fiber end will be required. Defining two additional terms will prove to be most convenient for further calculations. Therefore, for the fiber excitation occurring at the moment $t_E$ = 0, let us define $\mathrm{I}(t_E)$:
\begin{linenomath*}\begin{equation*}
\mathrm{I}(t_E)\equiv\int\limits_0^{t_E} E(t_\varepsilon)\mathrm{d}t_\varepsilon 
\tag{2.4}
\end{equation*}\end{linenomath*}
as a probability for photon emission prior to the moment $t_E$. The second useful term $\Lambda$:
\begin{linenomath*}\begin{equation*}
\Lambda\equiv\int\limits_{t_{min}}^{t_{max}} S(t_\pi)A(t_\pi)\mathrm{d}t_\pi
\tag{2.5}
\end{equation*}\end{linenomath*}
denotes the probability for emitted photon to actually reach the fiber end. With $t_{min}$ as minimal time required for the photon propagation\footnote{Minimal propagation time $t_{min}$ corresponds to the shortest distance path between the point of scintillation inside the fiber and the fiber end, i. e. to the photon emitted along the fiber axis.} and $t_{max}$ as maximal time of propagation permitted\footnote{Maximal propagation time $t_{max}$ is commonly considered to be defined by the optical condition for a total reflection off the fiber walls, i.e. by the critical angle for total reflection. However, this is not strictly true, which will be discussed in \autoref{app:chapC}.}, $\Lambda$ is governed by the probability for photon to become trapped by internal reflections inside the fiber and not to be absorbed within the fiber material.

For the $m$-th emitted photon to become the first to arrive at the fiber end, those $m-1$ previously emitted have to be lost, either by escaping the fiber or by absorption. Therefore, considering the photon combinations, factor $\alpha_m$ from (2.3):
\begin{linenomath*}\begin{equation*}
\alpha_m=\binom{n}{m-1}\left[\mathrm{I}(t_E)(1-\Lambda)\right]^{m-1}
\tag{2.6}
\end{equation*}\end{linenomath*}
is given by the probability for initial $m-1$ photons to be emitted prior to the emitting moment $t_E$ of the $m$-th one, and subsequently not to reach the fiber end. It is to be noted that $\alpha_m$ was constructed without arranging the lost photons in time, which is an approach validated in \autoref{app:chapA}. Furthermore, after being emitted at $t_E$, factor $\beta_m$:
\begin{linenomath*}\begin{equation*}
\beta_m=(n-m+1)E(t_E)A(t_P)S(t_P)
\tag{2.7}
\end{equation*}\end{linenomath*}
regulates the spatial direction, i.e. path length, and attenuation probability for the $m$-th photon. The number of remaining single photon selections is also taken into account. Finally, factor $\gamma_m$:
\begin{linenomath*}\begin{equation*}
\gamma_m=\left[1-\mathrm{I}(t_E)\right]^{n-m}
\tag{2.8}
\end{equation*}\end{linenomath*}
is restrained only by the requirement for the emission of remaining $n-m$ photons occurring after $t_E$, regardless of their outcome. Isolated combinatory factor is absent because all the selection options were depleted by $\alpha_m$ and $\beta_m$. It is to be noted that $\alpha_m$ and $\gamma_m$ are true probabilities, while $\beta_m$ is, in fact, a probability density.

With $\alpha_m$, $\beta_m$, $\gamma_m$ obtained, $p_m$ is completely determined by (2.3), and $f_n$, consequently, by (2.1). Therefore, writing $f_n$ explicitly:
\begin{linenomath*}\begin{align*}
\begin{split}
f_n=&E(t_E)A(t_P)S(t_P)\times\\
&\times\sum_{m=1}^n (n-m+1)\binom{n}{m-1}\left[\mathrm{I}(t_E)(1-\Lambda)\right]^{m-1}\left[1-\mathrm{I}(t_E)\right]^{n-m}
\end{split}
\tag{2.9}
\end{align*}\end{linenomath*}
it may be noted that by shifting a summation index a step backwards, a binomial expansion remains, reducing (2.9) into:
\begin{linenomath*}\begin{equation*}
f_n(t_P,t_E;l)=nE(t_E)A(t_P)S_l(t_P)\left[1-\mathrm{I}(t_E)\Lambda_l\right]^{n-1}
\tag{2.10}
\end{equation*}\end{linenomath*}
In (2.10) an explicit dependency on the position $l$ of a scintillation pulse origin along the fiber, the emission moment $t_E$ and the photon propagation time $t_P$ was written down for purposes of further calculations.

To complete the model, a number of emitted photons per scintillation pulse must be considered. For this a simple but effective and experimentally validated Poisson statistics is employed, defining the probability $P_N(n)$ for the emission of $n$ photons:
\begin{linenomath*}\begin{equation*}
P_N(n)=\frac{N^ne^{-N}}{n!}
\tag{2.11}
\end{equation*}\end{linenomath*}
parameterized only by their mean number $N$ per scintillation pulse. With this final distribution included, a probability density $\rho_l(t_P,t_E)$ for first arriving photon being assigned $t_E$ and $t_P$ may be obtained\footnote{The formal grounds for this step are addressed in \autoref{app:chapB}.}:
\begin{linenomath*}\begin{equation*}
\rho_l(t_P,t_E)=\sum_{n=1}^\infty P_N(n)f_n(t_P,t_E;l)
\tag{2.12}
\end{equation*}\end{linenomath*}
Entering (2.10) and (2.11) into (2.12), while shifting a summation index a step backwards:
\begin{linenomath*}\begin{equation*}
\rho_l(t_P,t_E)=Ne^{-N}E(t_E)A(t_P)S_l(t_P)\sum_{n=0}^\infty \frac{1}{n!}\left\{N[1-\mathrm{I}(t_E)\Lambda_l]\right\}^n
\tag{2.13}
\end{equation*}\end{linenomath*}
an exponential expansion may be recognized:
\begin{linenomath*}\begin{equation*}
\rho_l(t_P,t_E)=NE(t_E)A(t_P)S_l(t_P)e^{-N\mathrm{I}(t_E)\Lambda_l}
\tag{2.14}
\end{equation*}\end{linenomath*}
Furthermore, utilizing relation (2.2) yields a probability distribution $D_l(t_A)$ for the first photon arrival moment $t_A$:
\begin{linenomath*}\begin{equation*}
D_l(t_A)=\int\limits_{t_{min}(l)}^{t_{max}(l)} \rho_l(t_P;t_E=t_A-t_P)\mathrm{d}t_P
\tag{2.15}
\end{equation*}\end{linenomath*}
Finally, considering a probability to measure the first photon arrival time $t_A$ at one end of the fiber at a distance $l$ from scintillation pulse origin, with arrival time $t_A+\delta t$ at the other end distanced $L-l$, a coincidental timing distribution $R_l(\delta t)$ for a fiber of the length $L$ is found:
\begin{linenomath*}\begin{equation*}
R_l(\delta t)=\int\limits_0^\infty D_l(t_A)D_{L-l}(t_A+\delta t)\mathrm{d}t_A
\tag{2.16}
\end{equation*}\end{linenomath*}
From a symmetry in respect to the middle of the fiber, it is obvious that the following is true\footnote{This is actually only true if the spatial distribution $S$ of emitted photons is identical for both fiber ends, i.e. when $S(\theta)=S(\pi-\theta)$, with $S(\theta)$ being discussed in \autoref{sec:chap4}.}:
\begin{linenomath*}\begin{equation*}
R_l(\delta t)=R_{L-l}(-\delta t)
\tag{2.17}
\end{equation*}\end{linenomath*}

\section{Fine corrections}
\label{sec:chap3}
Let us consider the following cases:
\begin{enumerate}[1.)]
\item a photon is emitted at the moment $t_\varepsilon$ slightly prior to the first one reaching the fiber end ($t_\varepsilon<t_E$), but with a propagation time $t_\pi$ so long that its arrival follows or would follow that of the first incoming:
\begin{linenomath*}\begin{equation*}
t_\varepsilon+t_\pi>t_E+t_P
\tag{3.1}
\end{equation*}\end{linenomath*}
\item a photon is emitted slightly later than the first incoming ($t_\varepsilon>t_E$), but with a propagation time so short that, had it not been lost, its arrival at the fiber end would precede the first actual one:
\begin{linenomath*}\begin{equation*}
t_\varepsilon+t_\pi<t_E+t_P
\tag{3.2}
\end{equation*}\end{linenomath*}
\end{enumerate}
Evidently, the probability factors $\alpha_m$ and $\gamma_m$ given by (2.6) and (2.8), respectively, have to be corrected for such eventualities. Therefore, let us denote by $\Gamma_1$ the probability for an early-emitted photon to actually arrive late at the fiber end. Incorporating the condition (3.1) into integrals' limits and utilizing the relation (2.2), the defining expression for $\Gamma_1$ becomes:
\begin{linenomath*}\begin{align*}
\begin{split}
\Gamma_1(t_P,t_A;l)\equiv\int\limits_{t_A-t_{max}(l)}^{t_A-t_P} E(t_\varepsilon)\mathrm{d}t_\varepsilon&\int\limits_{t_A-t_\varepsilon}^{t_{max}(l)} S_l(t_\pi)A(t_\pi)\mathrm{d}t_\pi\\
&=\int\limits_{t_P}^{t_{max}(l)}  S_l(t_\pi)A(t_\pi)\mathrm{d}t_\pi\int\limits_{t_A-t_\pi}^{t_A-t_P} E(t_\varepsilon)\mathrm{d}t_\varepsilon
\end{split}
\tag{3.3}
\end{align*}\end{linenomath*}
Following the analogue procedure with the probability $\Gamma_2$ for a late-emitted photon to arrive prior to the defining moment $t_A$, the following remains:
\begin{linenomath*}\begin{align*}
\begin{split}
\Gamma_2(t_P,t_A;l)\equiv\int\limits_{t_A-t_P}^{t_A-t_{min}(l)} E(t_\varepsilon)\mathrm{d}t_\varepsilon&\int\limits_{t_{min}(l)}^{t_A-t_\varepsilon} S_l(t_\pi)A(t_\pi)\mathrm{d}t_\pi\\
&=\int\limits_{t_{min}(l)}^{t_P}  S_l(t_\pi)A(t_\pi)\mathrm{d}t_\pi\int\limits_{t_A-t_P}^{t_A-t_\pi} E(t_\varepsilon)\mathrm{d}t_\varepsilon
\end{split}
\tag{3.4}
\end{align*}\end{linenomath*}
The equality of two sides in both (3.3) and (3.4) is achieved by interchanging the integration order. Depending on the specific form of selected distributions, one of the sides may be greatly preferred for the actual numerical calculations.\\

\begin{figure}[h] 
\centering 
\includegraphics[width=0.5\textwidth,keepaspectratio]{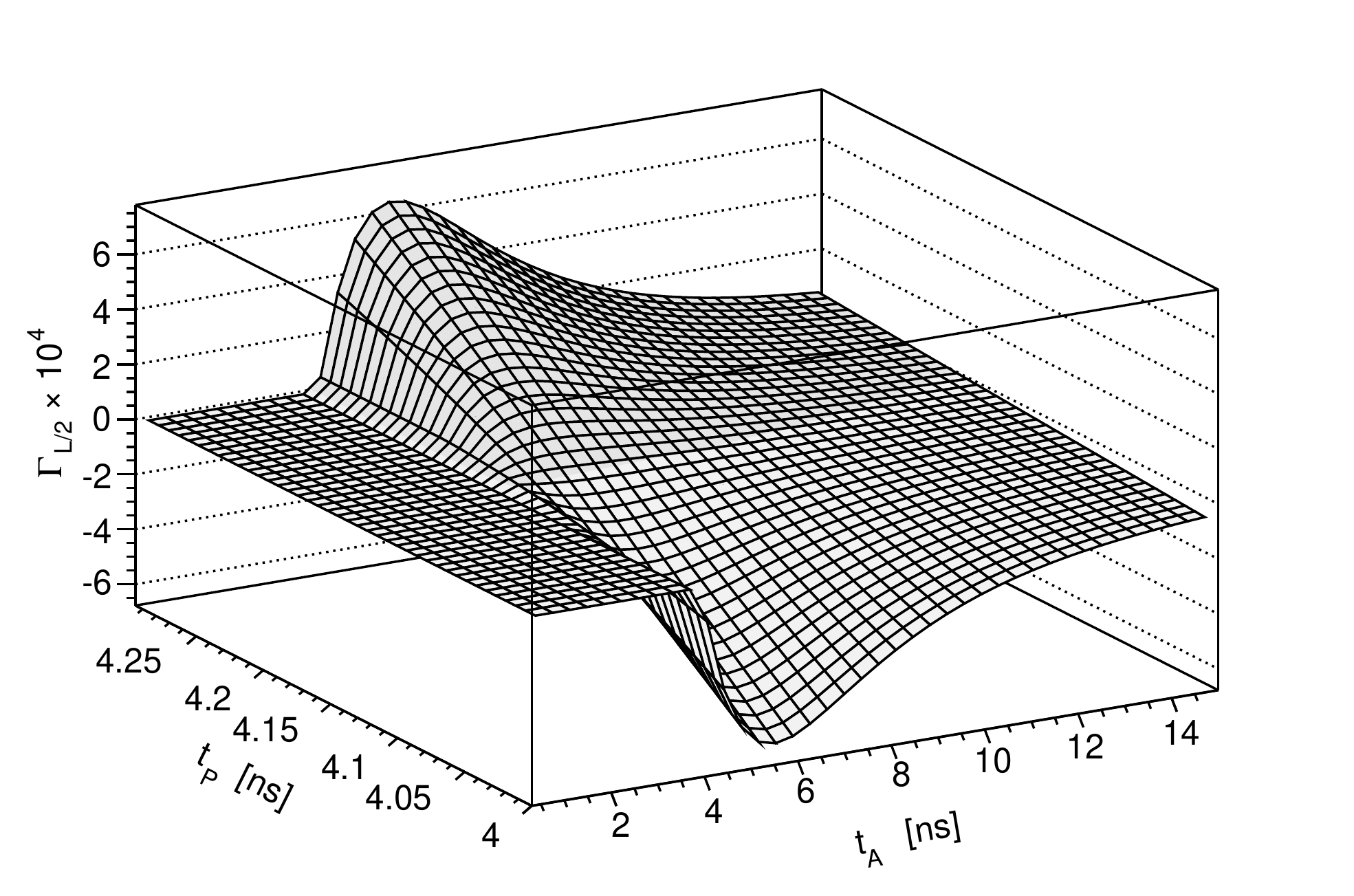}
\caption{An example of corrective probability $\Gamma_l(t_P,t_A)$ for $l=L/2$. The propagation time $t_P$ is limited by minimal and maximal values $t_{min}(l)$ and $t_{max}(l)$, respectively.}
\label{fig:fig2}
\end{figure}

Since the photons regulated by $\Gamma_1$ are, in fact, allowed to reach the fiber end, $\Gamma_1$ is to be added to the factor $\alpha_m$:
\begin{linenomath*}\begin{equation*}
\alpha_m=\binom{n}{m-1}\left[\mathrm{I}(t_E)(1-\Lambda_l)+\Gamma_1(t_P,t_A;l)\right]^{m-1}
\tag{3.5}
\end{equation*}\end{linenomath*}
while $\Gamma_2$ is to be subtracted from $\gamma_m$, not allowing any photon to precede the first incoming one:
\begin{linenomath*}\begin{equation*}
\gamma_m=\left[1-\mathrm{I}(t_E)-\Gamma_2(t_P,t_A;l)\right]^{n-m}
\tag{3.6}
\end{equation*}\end{linenomath*}
Following the steps leading from (2.1) to (2.14), it is invariably found that the corrected form for the probability density $\rho_l(t_P,t_E)$ equals:
\begin{linenomath*}\begin{equation*}
\rho_l(t_P,t_E)=NE(t_E)A(t_P)S_l(t_P)e^{-N\left[\mathrm{I}(t_E)\Lambda_l+\Gamma_l(t_P,t_A)\right]}
\tag{3.7}
\end{equation*}\end{linenomath*}
with a joined corrective probability $\Gamma_l(t_P,t_A)$ defined as:
\begin{linenomath*}\begin{equation*}
\Gamma_l(t_P,t_A)\equiv \Gamma_2(t_P,t_A;l)-\Gamma_1(t_P,t_A;l)
\tag{3.8}
\end{equation*}\end{linenomath*}
which, explicitly written, is of form:
\begin{linenomath*}\begin{align*}
\begin{split}
\Gamma_l(t_P,t_A)=\int\limits_{t_A-t_{max}(l)}^{t_A-t_{min}(l)} E(t_\varepsilon)\mathrm{d}t_\varepsilon&\int\limits_{t_{min}(l)}^{t_A-t_\varepsilon} S_l(t_\pi)A(t_\pi)\mathrm{d}t_\pi\\
&=\int\limits_{t_{min}(l)}^{t_{max}(l)} S_l(t_\pi)A(t_\pi)\mathrm{d}t_\pi\int\limits_{t_A-t_P}^{t_A-t_\pi} E(t_\varepsilon)\mathrm{d}t_\varepsilon
\end{split}
\tag{3.9}
\end{align*}\end{linenomath*}
An example of $\Gamma_{L/2}(t_P,t_A)$ -- as shown in \autoref{fig:fig2} -- was calculated for a specific selection of parameters and distributions introduced in \autoref{sec:chap4}. When applied to (2.16), the corrective effect from (3.7) would be suppressed by orders of magnitude within later presented Figures. Therefore, within \autoref{sec:chap4} and \autoref{sec:chap6} it was indeed more than sufficient to utilize (2.14) for numerical calculations.

\section{Model demonstration}
\label{sec:chap4}
To demonstrate the results of a proposed model, probability distributions with all the required parameters have to be defined. As the fiber prototype -- $L$ = 1.5 m in length -- a model BCF-10 from Saint-Gobain \cite{fib5} was selected. Due to the well established fact that the simple exponential description of the emission probability distribution $E(t_E)$ would be an oversimplification for the fast emitting plastic scintillators, a more adequate form was adopted from \cite{fib2,fib6}:
\begin{linenomath*}\begin{equation*}
E(t_E)=\frac{1}{r+1}\left(\frac{e^{-t_E/\tau_1}-e^{-t_E/\tau_2}}{\tau_1-\tau_2}+\frac{r}{\tau_3}e^{-t_E/\tau_3}\right)\Theta(t_E)
\tag{4.1}
\end{equation*}\end{linenomath*}
with a step function\footnote{Throughout the paper, the convention: $\Theta(t)=\begin{cases}0&\;\mathrm{if}\;t<0\\1&\;\mathrm{if}\;t\ge0\end{cases}$ \;\; will be used.} $\Theta(t_E)$ regulating causality. A decay time $\tau_1$ = 2.7 ns of the main fast component is reported in \cite{fib5}, while the values of additional parameters -- $\tau_2$ = 0.9 ns, $\tau_3$ = 14.2 ns and $r$ = 0.27 -- were selected from \cite{fib2,fib6}.\\

Path length dispersion $S(t_P)$ is deduced from the spatial distribution of emitted photons. Assuming isotropic scintillations, a normalized probability density $S(\theta)$ for photon being emitted under an angle $\theta$ relative to the fiber axis (\autoref{fig:fig1}) is given by:
\begin{linenomath*}\begin{equation*}
\mathrm{d}S(\theta)=-\tfrac{1}{2}\mathrm{d}(\cos\theta)
\tag{4.2}
\end{equation*}\end{linenomath*}
For the single-cladding fibers, light can be roughly considered to propagate only through the scintillating core, since that passing through the outermost cladding material is rapidly being lost due to the outer surface imperfections and detriments \cite{fib7}. Therefore, eliminating the contribution of the light refracted at the core-cladding interface, a photon propagation time $t_P(\theta;l)$ is given by the corresponding path length inside the fiber:
\begin{linenomath*}\begin{equation*}
t_P(\theta;l)=\frac{n_{core}l}{c}\frac{1}{\cos\theta}
\tag{4.3}
\end{equation*}\end{linenomath*}
with $n_{core}$ as a scintillating core refractive index and the speed of light $c$ in vacuum. Utilizing a relation (4.3), (4.2) is translated into:
\begin{linenomath*}\begin{equation*}
S_l(t_P)\mathrm{d}t_P=\frac{n_{core}l}{2c}\frac{\Theta\left[t_P-t_{min}(l)\right]\Theta\left[t_{max}(l)-t_P\right]}{t_P^2}\mathrm{d}t_P
\tag{4.4}
\end{equation*}\end{linenomath*}
where $t_{min}(l)$ is a minimal time required for the light propagation\; ($\theta=0$):
\begin{linenomath*}\begin{equation*}
t_{min}(l)=\frac{n_{core}l}{c}
\tag{4.5}
\end{equation*}\end{linenomath*}
A particular value of $n_{core}$ = 1.60 is obtained from \cite{fib5}. With a cladding refractive index of $n_{clad}$ = 1.49, $t_{max}(l)$ is a maximal propagation time restricted by the critical angle for a total internal reflection off the core-cladding interface \qquad\qquad ($\cos\theta_{critical}=n_{clad}/n_{core}$):
\begin{linenomath*}\begin{equation*}
t_{max}(l)=\frac{n_{core}^2l}{n_{clad}c}
\tag{4.6}
\end{equation*}\end{linenomath*}\\

\begin{figure}[t] 
\centering 
\includegraphics[width=0.5\textwidth,keepaspectratio]{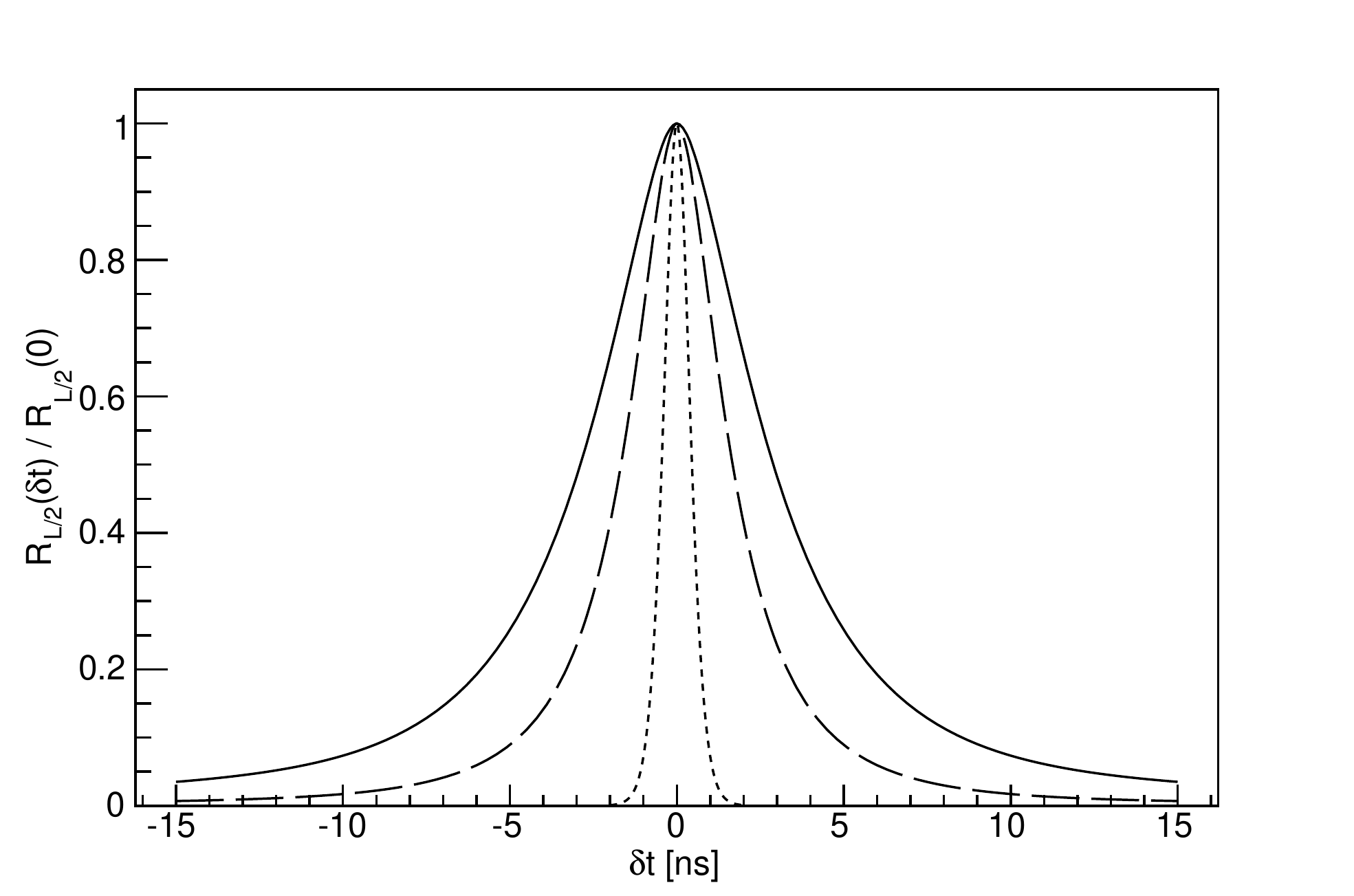}
\caption{Rescaled coincidental timing distributions for $l=L/2$ and the average of $N$ = $10^1$ (full line), $10^2$ (long-dashed line), $10^3$ (short-dashed line) photons. The improvement in resolution for an increased $N$ is evident.}
\label{fig:fig3}
\end{figure}

\begin{figure}[h!] 
\centering 
\begin{overpic}[scale=0.455]{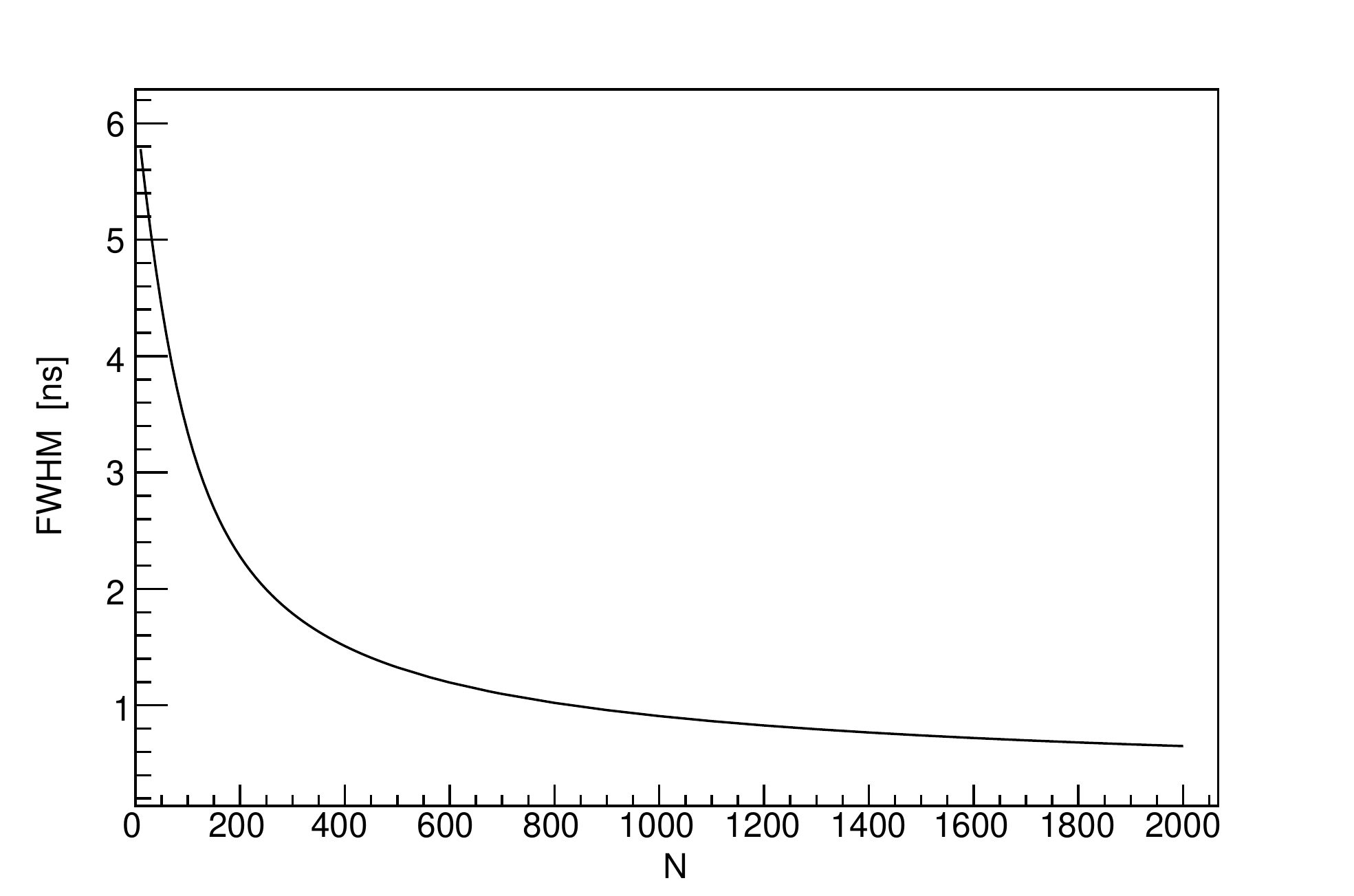}
\put(27,18){\includegraphics[scale=0.28]{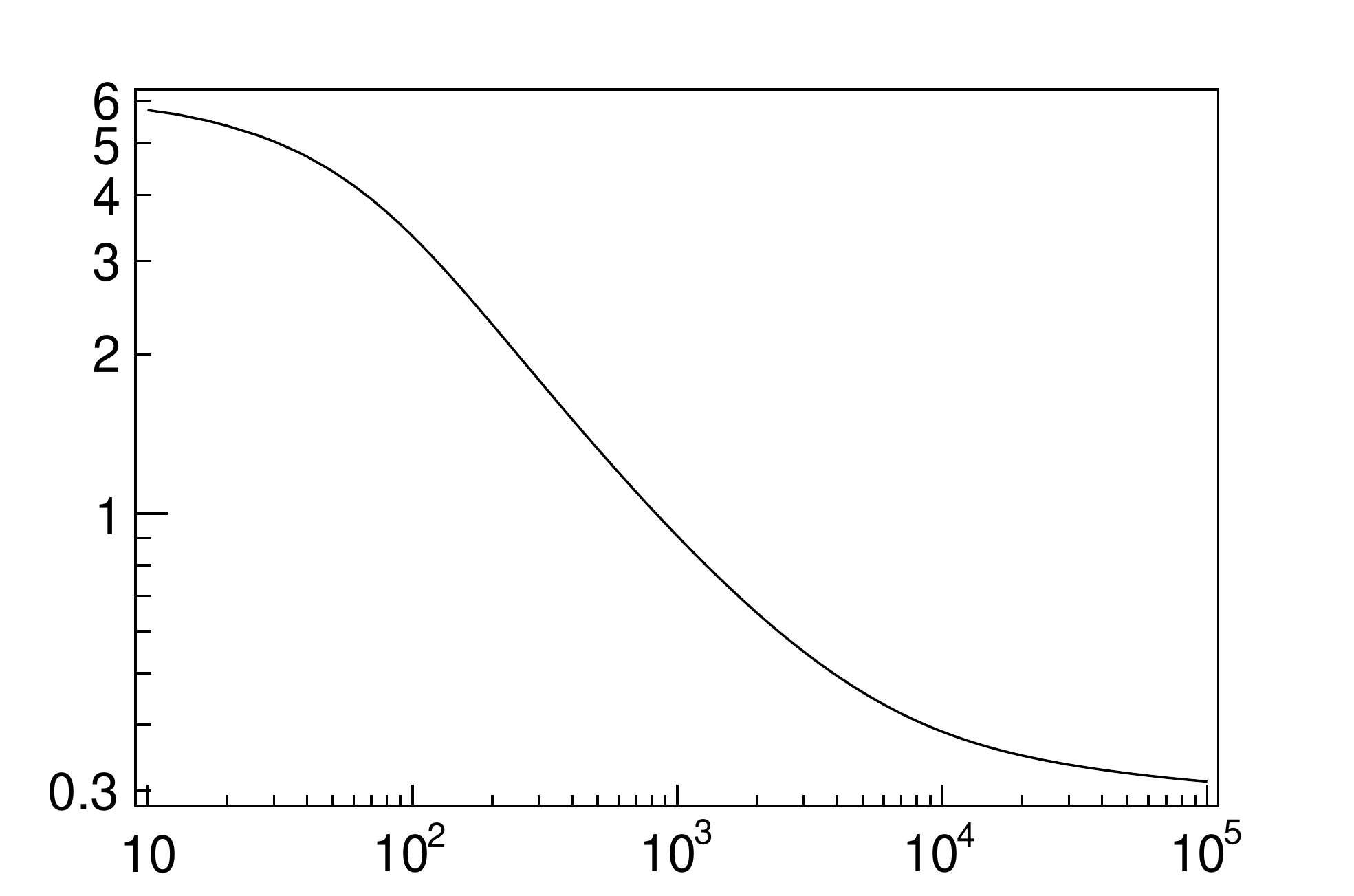}}
\end{overpic}
\caption{FWHM dependency on the mean $N$ of emitted photons. The isolated frame shows the plot with logarithmic scale for both axes, covering the wider range for $N$. Note that FWHM($N=10^5$) $>$ 0.3 ns.}
\label{fig:fig4}
\end{figure}

\begin{figure}[h!] 
\centering 
\includegraphics[width=0.5\textwidth,keepaspectratio]{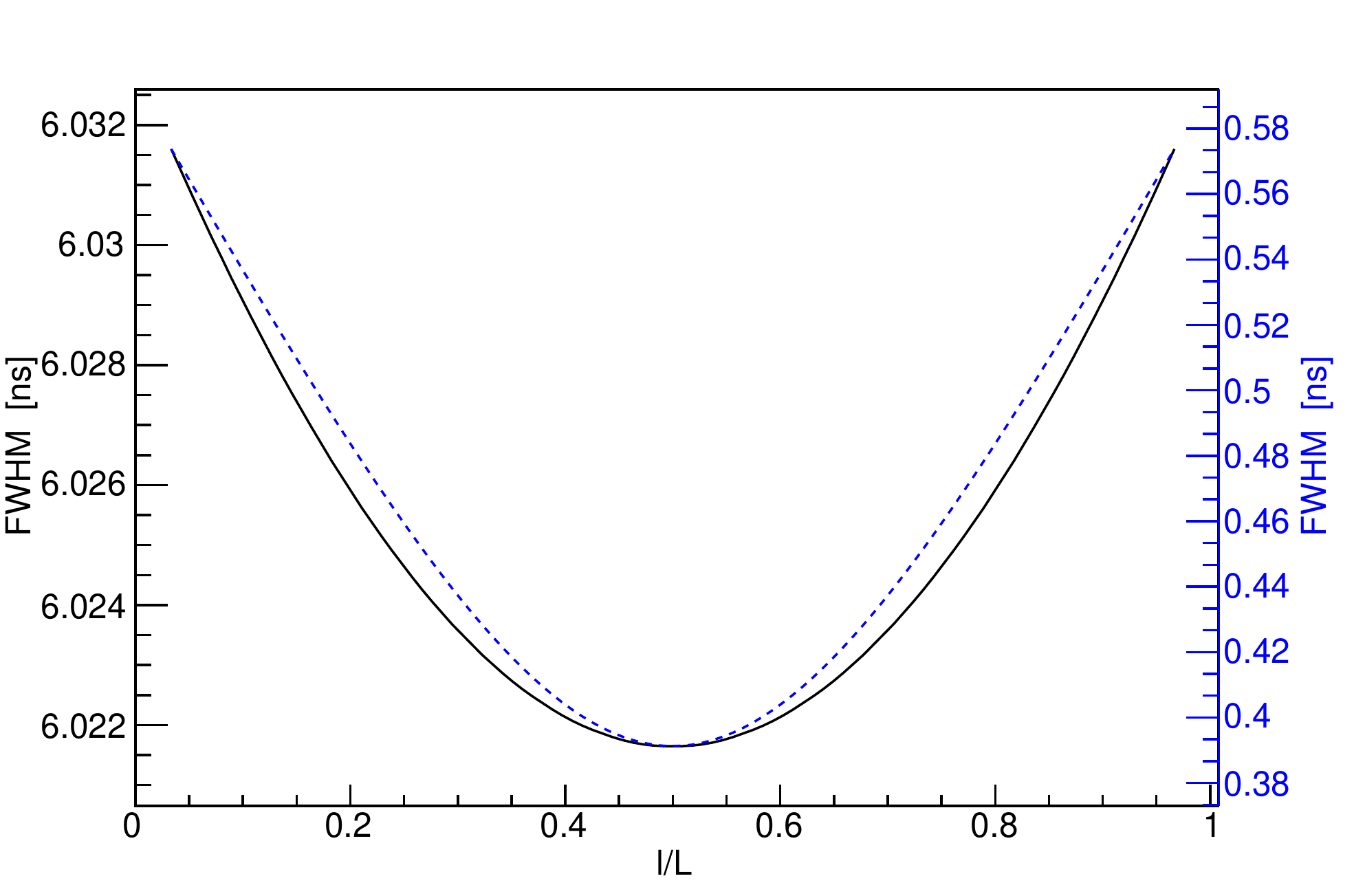}
\caption{FWHM value dependent on the position of a scintillating pulses' point of origin. Full line (left scale) shows the case for  $N$ = 4, while dashed one (right scale) for $N=10^4$.}
\label{fig:fig5}
\end{figure}

Finally, for an attenuation factor a simple exponential dependency on the path length $A(\theta)=e^{-l/(\lambda\cos\theta)}$ was adopted, with a single value $\lambda$ = 2.2 m for the attenuation length available \cite{fib5}. Utilizing (4.3), $A(\theta)$ translates into:
\begin{linenomath*}\begin{equation*}
A(t_P)=e^{-ct_P/(n_{core}\lambda)}
\tag{4.7}
\end{equation*}\end{linenomath*}\\

\autoref{fig:fig3} shows the rescaled coincidental timing distributions for $l=L/2$ and the mean number of $N$ = $10^1$, $10^2$, $10^3$ photons, emphasizing the decrease in resolution defining FWHM value. A fully expected improvement in resolution for the increased mean number $N$ of emitted photons is explicitly presented in \autoref{fig:fig4}. Though, purely mathematically, it is also to be expected that FWHM drops to 0 with $N$ approaching infinity, even for the excessive number of $10^5$ photons, FWHM amounts to more than 0.3 ns. Additionally, for $N=4$ and $N=10^4$ \autoref{fig:fig5} shows the influence of the scintillating pulses' position along the fiber upon the attainable FWHM values.

\begin{figure*}[t]
\centering
\begin{tabular}{cc}
\epsfig{file=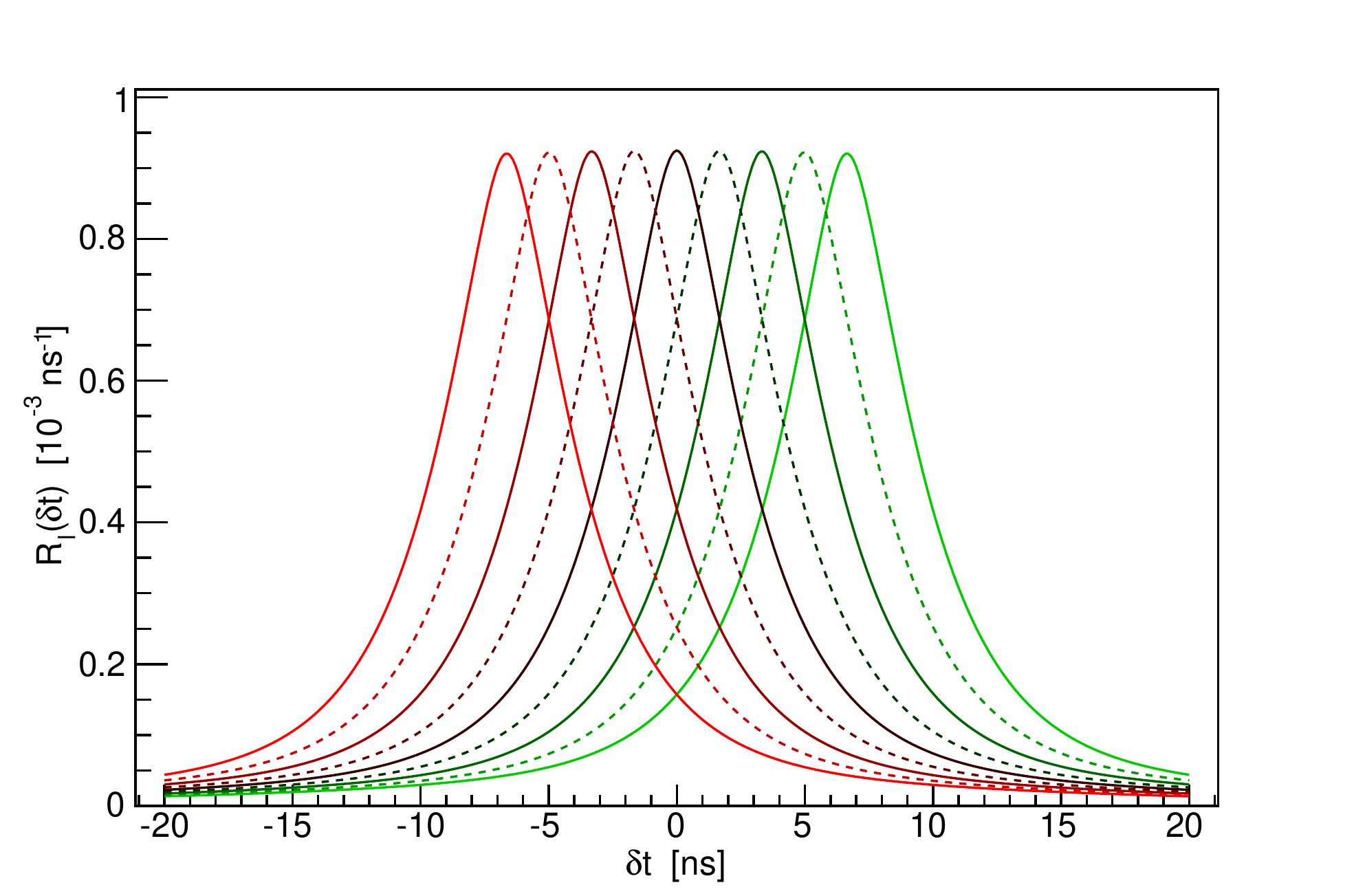,width=0.5\textwidth,keepaspectratio,clip=} &
\epsfig{file=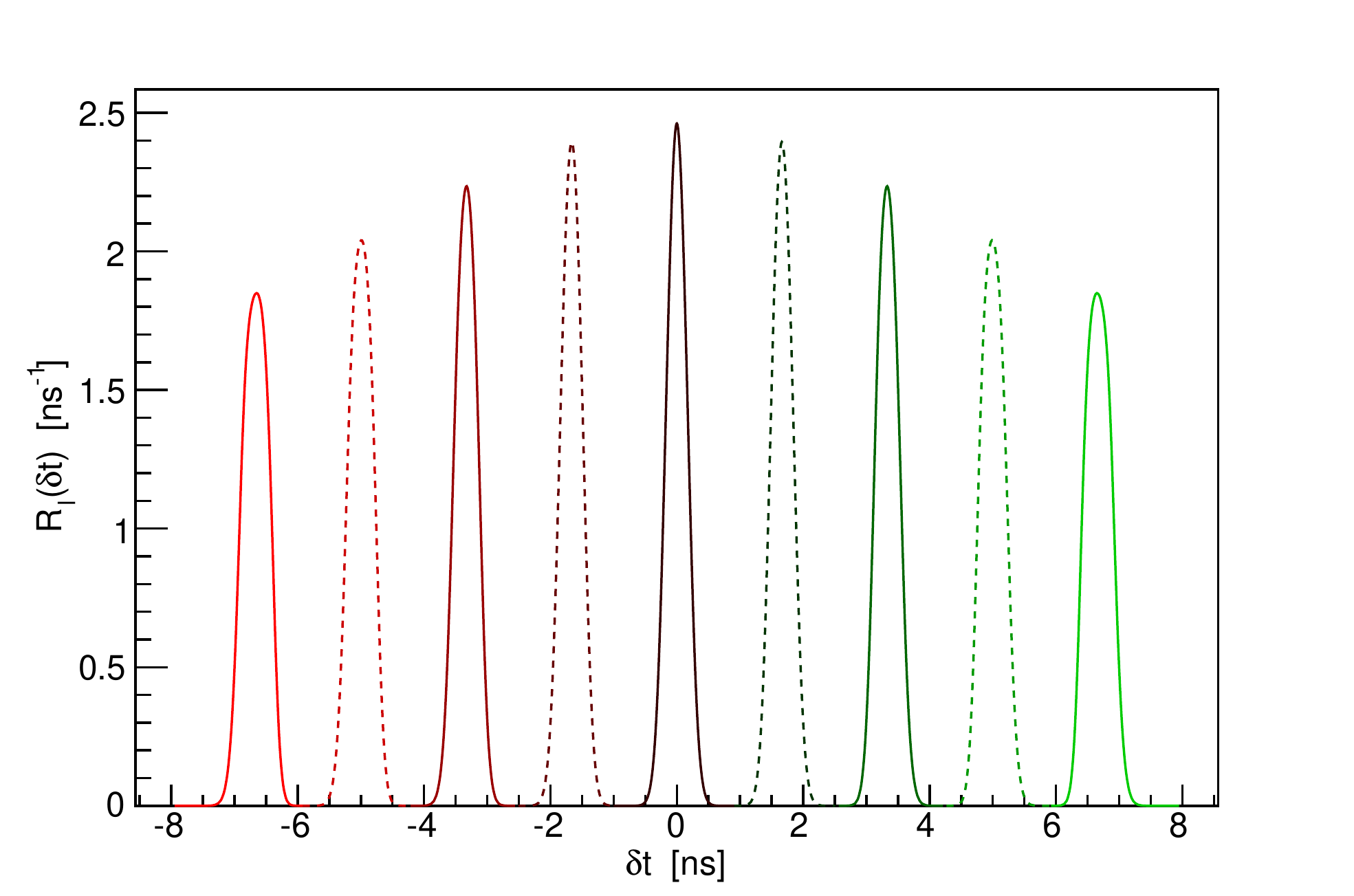,width=0.5\textwidth,keepaspectratio,clip=}
\end{tabular}
\caption{Coincidental timing distributions for $N$ = 4 (left) and $N=10^4$ (right). For a given $N$, 9 equidistant positions of scintillating pulses along the fiber length (from right to left) are given by: $l/L=0.1i$\; ($i=1,\ldots,9$).}
\label{fig:fig6}
\end{figure*}

In \autoref{fig:fig6} the coincidental timing distributions are given for 9 equidistant pulses' origins and separate cases of $N=4$ and $N=10^4$ photons.

\section{Photon count}
\label{sec:chap5}
Most relevant for the experimental purposes, physical quantities such as the photon detection efficiency and the mean number of detected photons deserve a special consideration. For a proposed model (2.16), a photon detection efficiency $\Psi_N(l)$ is contained within the area below the coincidental timing distribution curve:
\begin{linenomath*}\begin{equation*}
\Psi_N(l)=\int\limits_{-\infty}^{\infty} R_l(\delta t)\mathrm{d}(\delta t)
\tag{5.1}
\end{equation*}\end{linenomath*}
However, due to the complexity of (2.16), little could be achieved with the approach from (5.1). Fortunately, there is an alternate procedure yielding the exact solution for $\Psi_N(l)$. Let us, therefore, consider the probability $\eta_n(l)$ for at least one photon of total $n$ emitted to reach every end of the fiber:
\begin{linenomath*}\begin{equation*}
\eta_n(l)=\sum_{m=1}^{n-1}\sum_{k=1}^{n-m} \binom{n}{m}\binom{n-m}{k}\Lambda_l^m\Lambda_{L-l}^k(1-\Lambda_l-\Lambda_{L-l})^{n-m-k}
\tag{5.2}
\end{equation*}\end{linenomath*}
Evidently, it is given by all the available combinations of $m$ photons reaching one end, $k$ photons the other, with the remainder of $n-m-k$ being lost. It may be readily shown that $\eta_n(l)$ equals:
\begin{linenomath*}\begin{equation*}
\eta_n(l)=1-(1-\Lambda_l)^n-(1-\Lambda_{L-l})^n+(1-\Lambda_l-\Lambda_{L-l})^n
\tag{5.3}
\end{equation*}\end{linenomath*}
The transparency of previous result may be appreciated, as it excludes the probability for all photons not to arrive at one of the ends, while correcting for a double intake of none of them reaching either end. This exact form enables a straightforward calculation of $\Psi_N(l)$, when the probability for a particular number of emitted photons is considered:
\begin{linenomath*}\begin{equation*}
\Psi_N(l)=\sum_{n=2}^{\infty} P_N(n)\eta_n(l)
\tag{5.4}
\end{equation*}\end{linenomath*}
It is to be noted that (5.4) requires at least two photons, one for each fiber end. Prior to disclosing the solution for (5.4), we propose an alternate starting point for calculation of $\Psi_N(l)$:
\begin{linenomath*}\begin{align*}
\begin{split}
\Psi_N(l)=\sum_{n=2}^{\infty}\sum_{m=1}^{n-1}&\left[P_N(m)\sum_{k=1}^m \binom{m}{k}\Lambda_l^k(1-\Lambda_l)^{m-k}\right]\times\\
&\times\left[P_N(n-m)\sum_{k=1}^{n-m} \binom{n-m}{k}\Lambda_{L-l}^k(1-\Lambda_{L-l})^{n-m-k}\right]
\end{split}
\tag{5.5}
\end{align*}\end{linenomath*}
based upon the approach discussed in \autoref{app:chapB} and considering the arrival of photons at each fiber end separately.\\\\Both (5.4) and (5.5) consistently yield:
\begin{linenomath*}\begin{equation*}
\Psi_N(l)=1-e^{-N\Lambda_l}-e^{-N\Lambda_{L-l}}+e^{-N(\Lambda_l+\Lambda_{L-l})}
\tag{5.6}
\end{equation*}\end{linenomath*}\\

\begin{figure}[h!] 
\centering 
\includegraphics[width=0.5\textwidth,keepaspectratio]{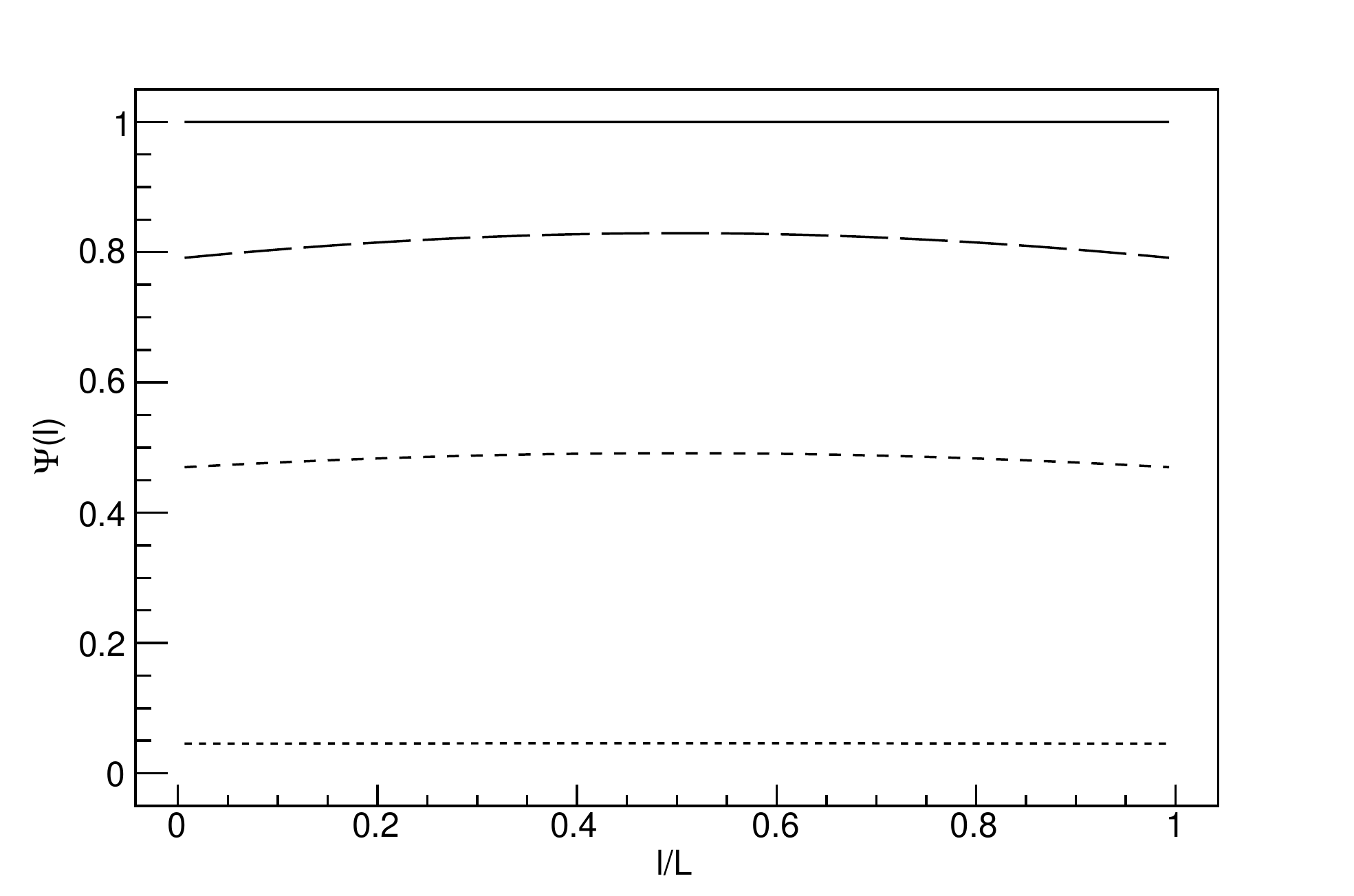}
\caption{Photon detection efficiencies for   $N$ = 10 (short-dashed line),\qquad\qquad 50 (medium-dashed line), 100 (long-dashed line), 500 (full line) photons. The placement within absolute [0,1] probability range is emphasized.}
\label{fig:fig7}
\end{figure}

\begin{figure}[t] 
\centering 
\begin{overpic}[scale=0.455]{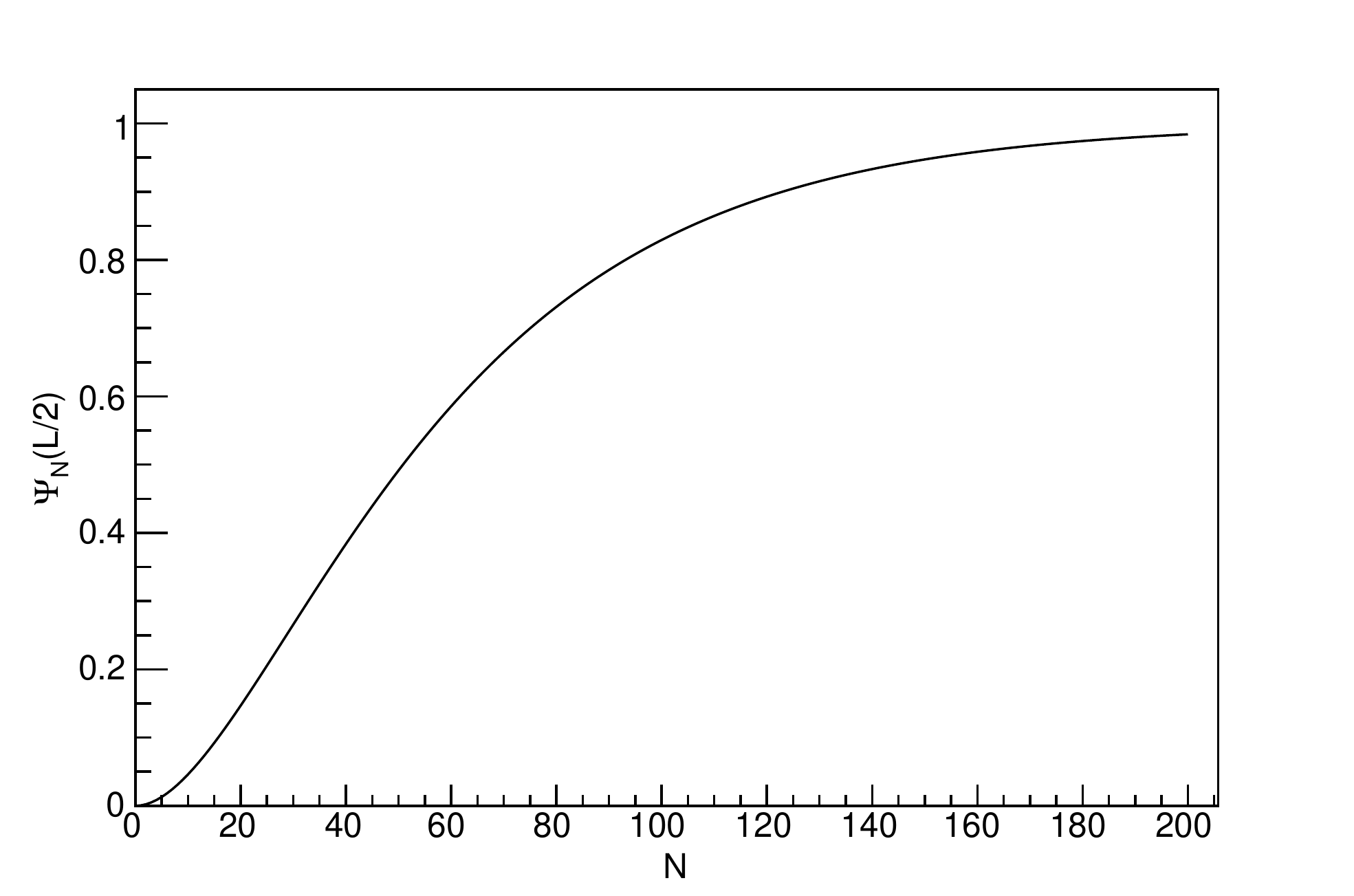}
\put(39,8){\includegraphics[scale=0.23]{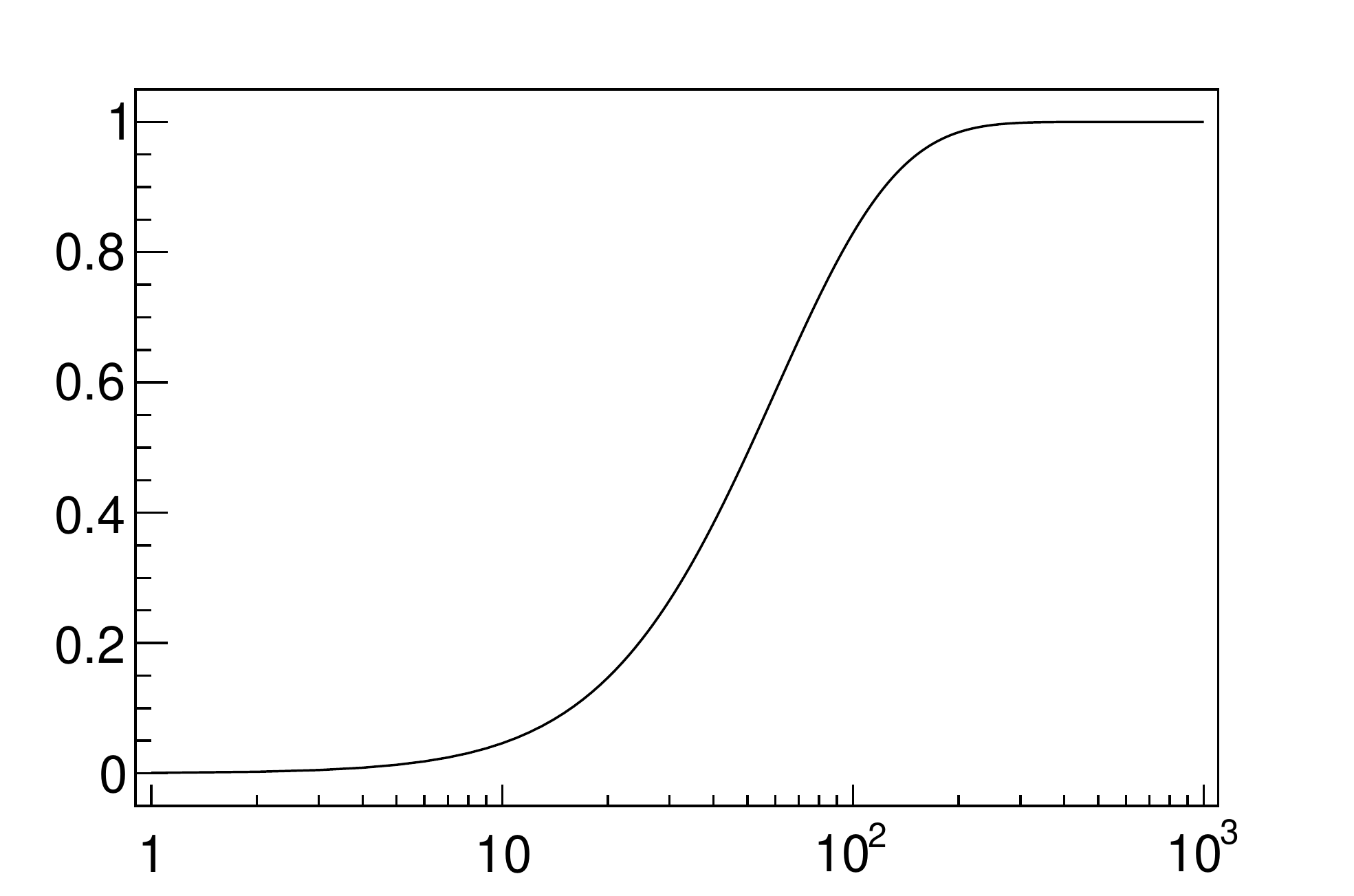}}
\end{overpic}
\caption{A photon detection efficiency for $l=L/2$ and a varying mean of the photons emitted. An isolated frame presents a logarithmic scale for $N$, encompassing a wider range of values while revealing a probability saturation for large $N$.}
\label{fig:fig8}
\end{figure}

\autoref{fig:fig7} plots the solution (5.6) for $N$ = 10, 50, 100, 500 photons, emphasizing the placement within absolute probability scale [0,1]. \autoref{fig:fig8} shows $\Psi_N(L/2)$ dependent on average number of photons emitted. There is an additional observation to be made, based on the fact that for a sufficiently large $N$ the detection efficiency saturates around 1. It is implied that a limiting case of the coincidental timing distribution corresponds with a representation of a delta-distribution:
\begin{linenomath*}\begin{equation*}
\lim_{N\rightarrow\infty} R_{l}(\delta t;N)=\delta\left[\delta t+t_{min}(l)-t_{min}(L-l)\right]
\tag{5.7}
\end{equation*}\end{linenomath*}
encompassing, in fact, an infinite set of representations, one for every single arbitrary selection of distributions and parameters involved.\\

Besides the efficiency for detecting any photons at both fiber ends, their mean number may also be deduced. For measurements involving the coincidental timing resolution, a coincidence of photon arrivals at both ends is a prerequisite that defines the starting point for the calculation of a coincidental average $\mathrm{M}_N(l)$:
\begin{linenomath*}\begin{align*}
\begin{split}
\mathrm{M}_N(l)=\frac{1}{\Psi_N(l)}&\sum_{n=2}^{\infty}\sum_{m=1}^{n-1}\left[P_N(m)\sum_{k=1}^m k\binom{m}{k}\Lambda_l^k(1-\Lambda_l)^{m-k}\right]\times\\
&\times\left[P_N(n-m)\sum_{k=1}^{n-m} \binom{n-m}{k}\Lambda_{L-l}^k(1-\Lambda_{L-l})^{n-m-k}\right]
\end{split}
\tag{5.8}
\end{align*}\end{linenomath*}
Due to the coincidence requirement, normalization for the event detecting probability less than 1 must be performed, yielding a simple solution:
\begin{linenomath*}\begin{equation*}
\mathrm{M}_N(l)=\frac{N\Lambda_l}{1-e^{-N\Lambda_l}}
\tag{5.9}
\end{equation*}\end{linenomath*}
The fact that (5.9) is devoid of a term $\Lambda_{L-l}$ is consistent with an independence of the fiber ends, formally supported by (B.2) (\autoref{app:chapB}).

\section{Triggering feature}
\label{sec:chap6}
Photon resolving devices offer an experimental possibility for testing the various aspects of the single-photon and/or multi-photon statistics. In \cite{fib3} it is experimentally demonstrated that by eliminating the events with low number of detected photons (triggering) at either fiber end, the width of the remaining coincidental timing distribution significantly decreases, i.e. resolution improves. Therefore, our goal is to include such triggering option within the theoretical model.\\

Let us demand that only the events with more than $\Delta$ detected photons are acceptable from one of the fiber ends. Hence, the probability $\pi_{n,\Delta}(l)$ for at least $\Delta$ out of $n$ photons reaching the end is required:
\begin{linenomath*}\begin{equation*}
\pi_{n,\Delta}(l)=\Theta(n-\Delta)\sum_{k=\Delta}^n \binom{n}{k}\Lambda_l^k(1-\Lambda_l)^{n-k}
\tag{6.1}
\end{equation*}\end{linenomath*}
Furthermore, let us consider an effect on a factor $p_m$ from (2.3), which labeling will be extended to $p_{n,m}(l)$ for further use. Since it is assumed within $p_{n,m}(l)$ that initial $m-1$ of $n$ photons were lost, while $m$-th one does reach the fiber end, $n-m$ photons remain for the probability $\pi$ to regulate:
\begin{linenomath*}\begin{equation*}
p_{n,m}(l)\qquad\longrightarrow\qquad p_{n,m,\Delta}(l)=p_{n,m}(l)\pi_{n-m,\Delta}(l)
\tag{6.2}
\end{equation*}\end{linenomath*}
Finally, a probability $f_n$ from (2.1) must be corrected for allowing only the acceptance of more than $\Delta$ photons:
\begin{linenomath*}\begin{equation*}
f_{n,\Delta}(l)=\Theta(n-\Delta-1)\sum_{m=1}^n p_{n,m,\Delta}(l)
\tag{6.3}
\end{equation*}\end{linenomath*}
Upon entering (6.3) into (2.12), the following expression remains:
\begin{linenomath*}\begin{equation*}
\rho_l(t_P,t_E;\Delta)=\sum_{n=\Delta+1}^{\infty} P_N(n)\sum_{m=1}^{n-\Delta} p_{n,m}(l)\sum_{k=\Delta}^{n-m} \binom{n-m}{k}\Lambda_l^k(1-\Lambda_l)^{n-m-k}
\tag{6.4}
\end{equation*}\end{linenomath*}
with a triggering parameter $\Delta$ appearing inside the limits of all the sums. Evidently, for $\Delta$ = 0 (6.4) reduces to (2.14). Since trigger values may be separately set for every fiber end, the final coincidental timing model incorporates both of them:
\begin{linenomath*}\begin{equation*}
R_l(\delta t;\Delta_1,\Delta_2)=\int\limits_0^{\infty} D_l(t_A;\Delta_1)D_{L-l}(t_A+\delta t;\Delta_2)\mathrm{d}t_A
\tag{6.5}
\end{equation*}\end{linenomath*}

\begin{figure}[t] 
\centering 
\includegraphics[width=0.5\textwidth,keepaspectratio]{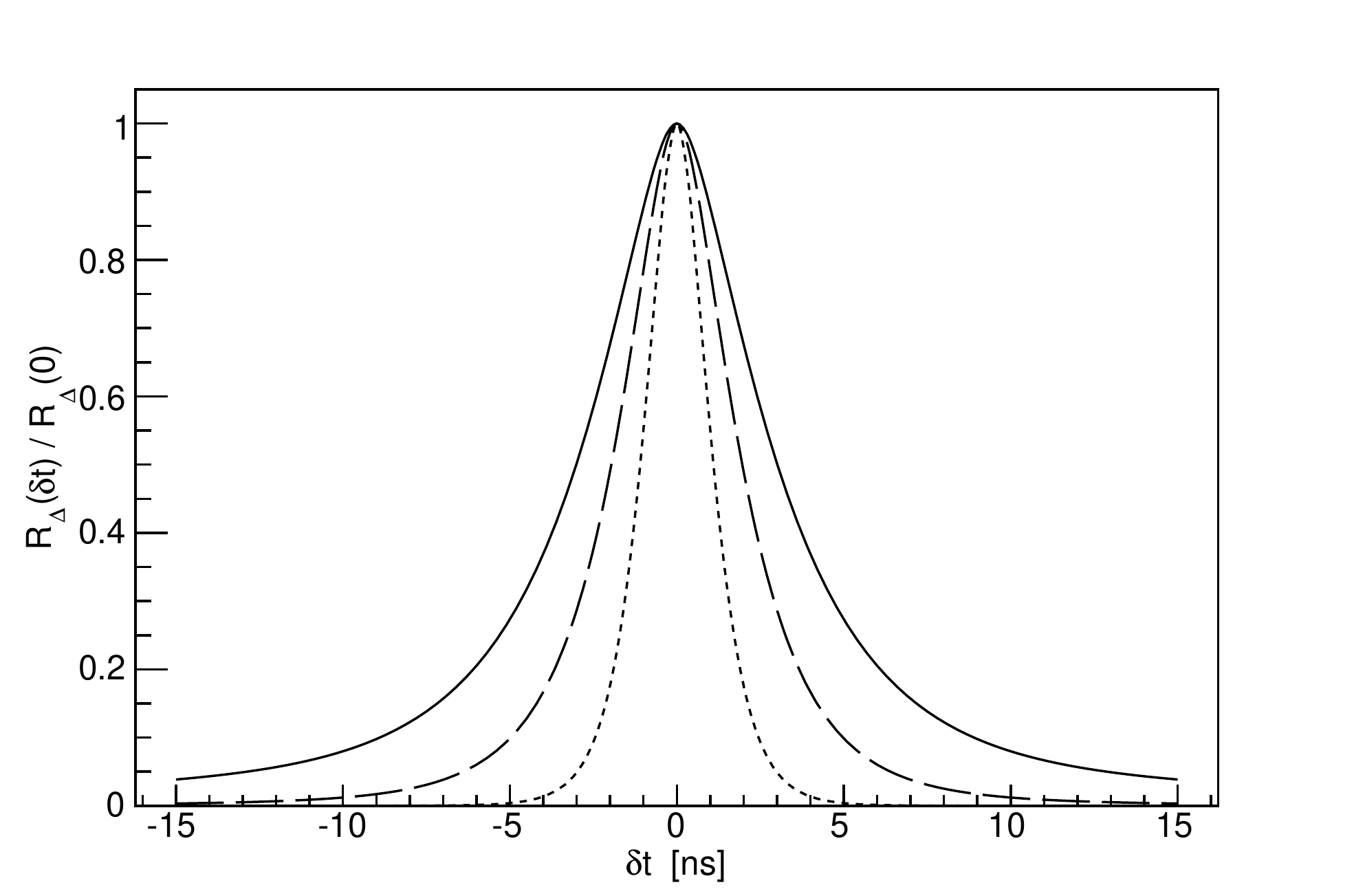}
\caption{Rescaled coincidental timing distributions for $l=L/2$ and $N$ = 4, showing the improvement in resolution due to triggering. Trigger values were set to be equal: $\Delta_1=\Delta_2=\Delta$, with $\Delta$ = 0 (full line), 1 (long-dashed line), 4 (short-dashed line).}
\label{fig:fig9}
\end{figure}

\begin{figure}[t] 
\centering 
\begin{overpic}[scale=0.455]{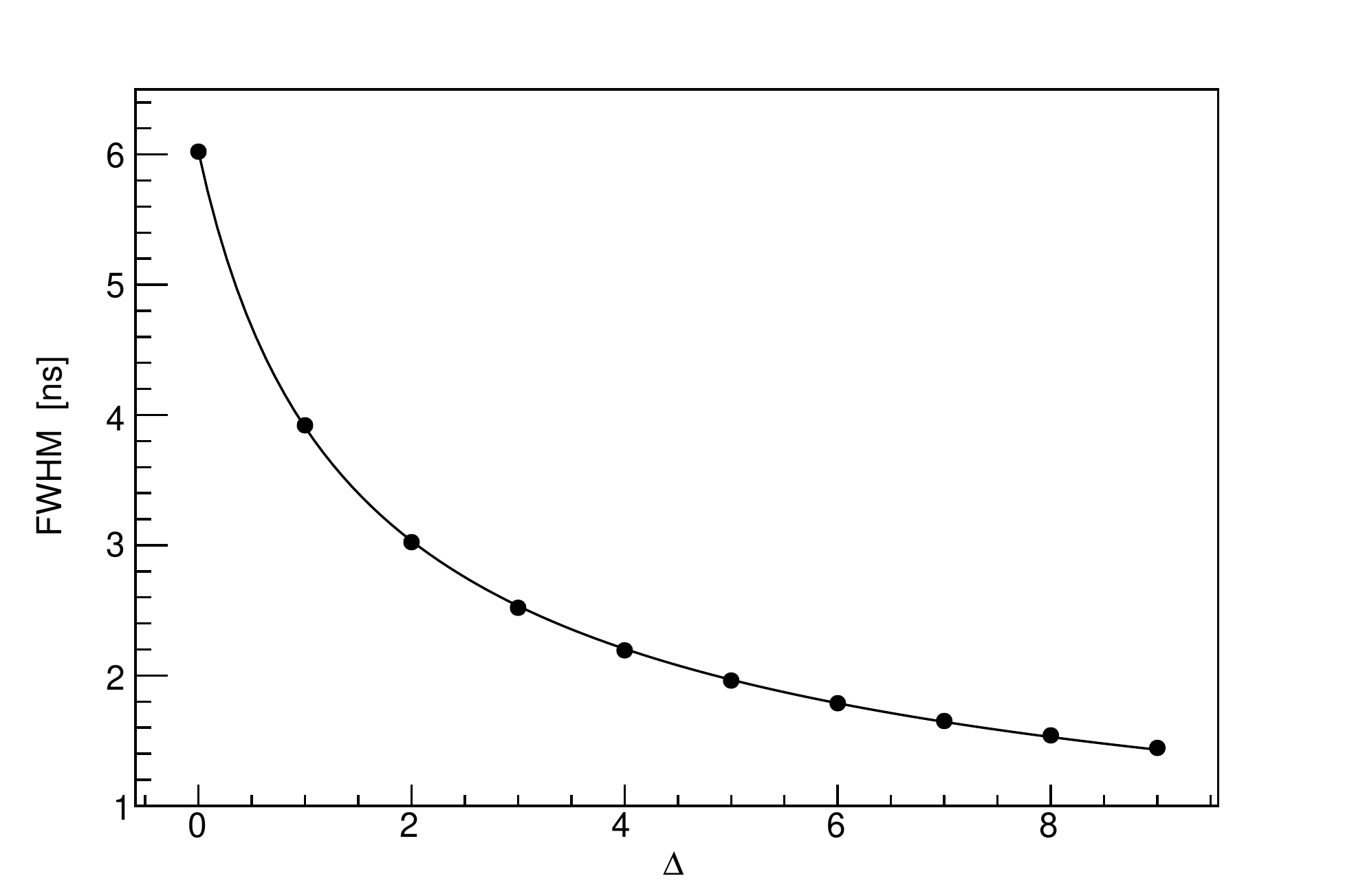}
\put(37,23){\includegraphics[scale=0.24]{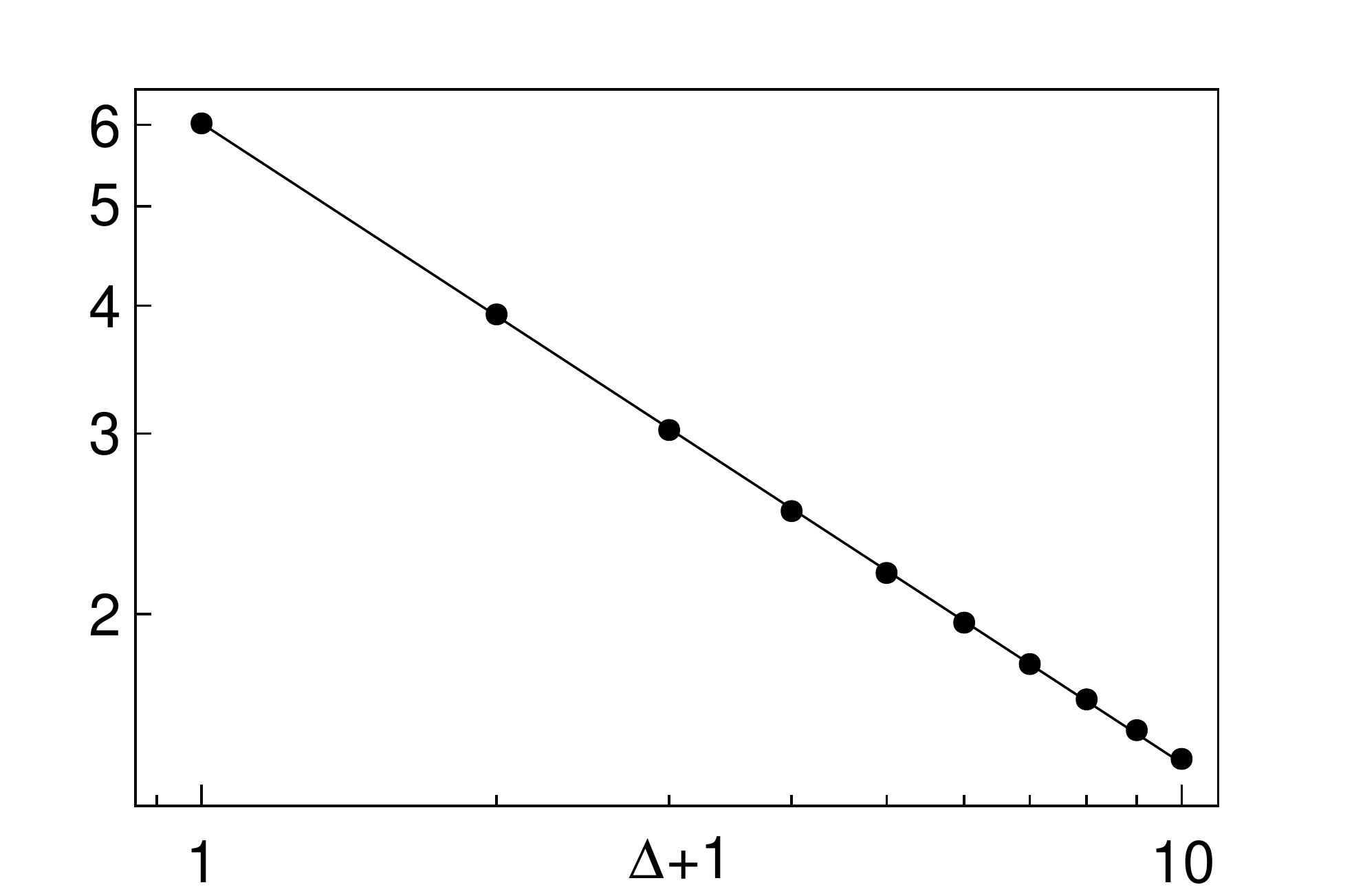}}
\end{overpic}
\caption{Triggering induced decrease in FWHM for $l=L/2$ and $N=4$, with $\Delta_1=\Delta_2=\Delta$. An isolated frame presents both scales as logarithmic, emphasizing the reasonable power-law trend, shown by the full line \qquad $\mathrm{FWHM}\propto(\Delta+1)^{-0.624}$.}
\label{fig:fig10}
\end{figure}

\autoref{fig:fig9} shows the effects of triggering for the case of $N=4$ photons with $\Delta_1=\Delta_2=$ 0, 1, 4, supporting the resolution improvement observed in \cite{fib3}. Moreover, \autoref{fig:fig10} presents the reduce in FWHM for $N$ = 4 and few values of $\Delta_1=\Delta_2=\Delta$. It was found that points from \autoref{fig:fig10} follow the trend $\mathrm{FWHM}\propto(\Delta+1)^{-0.624}$ reasonably well.\\

The triggering generalization is also easily extended to the model for a photon detection efficiency and a mean number of detected photons. Therefore, a trigger inclusive form for the photon detection efficiency equals:
\begin{linenomath*}\begin{align*}
\begin{split}
\Psi_N(l;\Delta_1,\Delta_2)=&\sum_{n=2+\Delta_1+\Delta_2}^{\infty}\sum_{m=1+\Delta_1}^{n-1-\Delta_2}\left[P_N(m)\sum_{k=1+\Delta_1}^m \binom{m}{k}\Lambda_l^k(1-\Lambda_l)^{m-k}\right]\times\\
&\times\left[P_N(n-m)\sum_{k=1+\Delta_2}^{n-m} \binom{n-m}{k}\Lambda_{L-l}^k(1-\Lambda_{L-l})^{n-m-k}\right]
\end{split}
\tag{6.6}
\end{align*}\end{linenomath*}
while that for the coincidental mean of detected photons:
\begin{linenomath*}\begin{align*}
\begin{split}
\mathrm{M}_N&(l;\Delta_1,\Delta_2)=\frac{1}{\Psi_N(l;\Delta_1,\Delta_2)}\times\\
&\times\sum_{n=2+\Delta_1+\Delta_2}^{\infty}\sum_{m=1+\Delta_1}^{n-1-\Delta_2}\left[P_N(m)\sum_{k=1+\Delta_1}^m k\binom{m}{k}\Lambda_l^k(1-\Lambda_l)^{m-k}\right]\times\\
&\qquad\qquad\times\left[P_N(n-m)\sum_{k=1+\Delta_2}^{n-m} \binom{n-m}{k}\Lambda_{L-l}^k(1-\Lambda_{L-l})^{n-m-k}\right]
\end{split}
\tag{6.7}
\end{align*}\end{linenomath*}
By setting all the available trigger values within (6.5), (6.6) and/or (6.7) to 0, a model without the triggering effects remains. Entering $\Delta_2=-1$ into (6.7) leaves a non-coincidental average of photons detected at the single fiber end. For experiments considering not only the photon-inclusive events, but also the pedestal measurement, $\Delta_1=\Delta_2=-1$ may be inserted, simplifying the photon average calculation:
\begin{linenomath*}\begin{equation*}
\mathrm{M}_N(l;\Delta_1=-1,\Delta_2=-1)=N\Lambda_l
\tag{6.8}
\end{equation*}\end{linenomath*}
However, in a course of actual measurements, a single photon detection efficiency for the photon resolving detectors must be considered. \autoref{app:chapC} addresses this subject, essential for the correct estimation of the emitted photons average $N$.

\section{Experimental verification}
\label{sec:chap7}
\begin{figure}[t] 
\centering 
\includegraphics[width=0.5\textwidth,keepaspectratio]{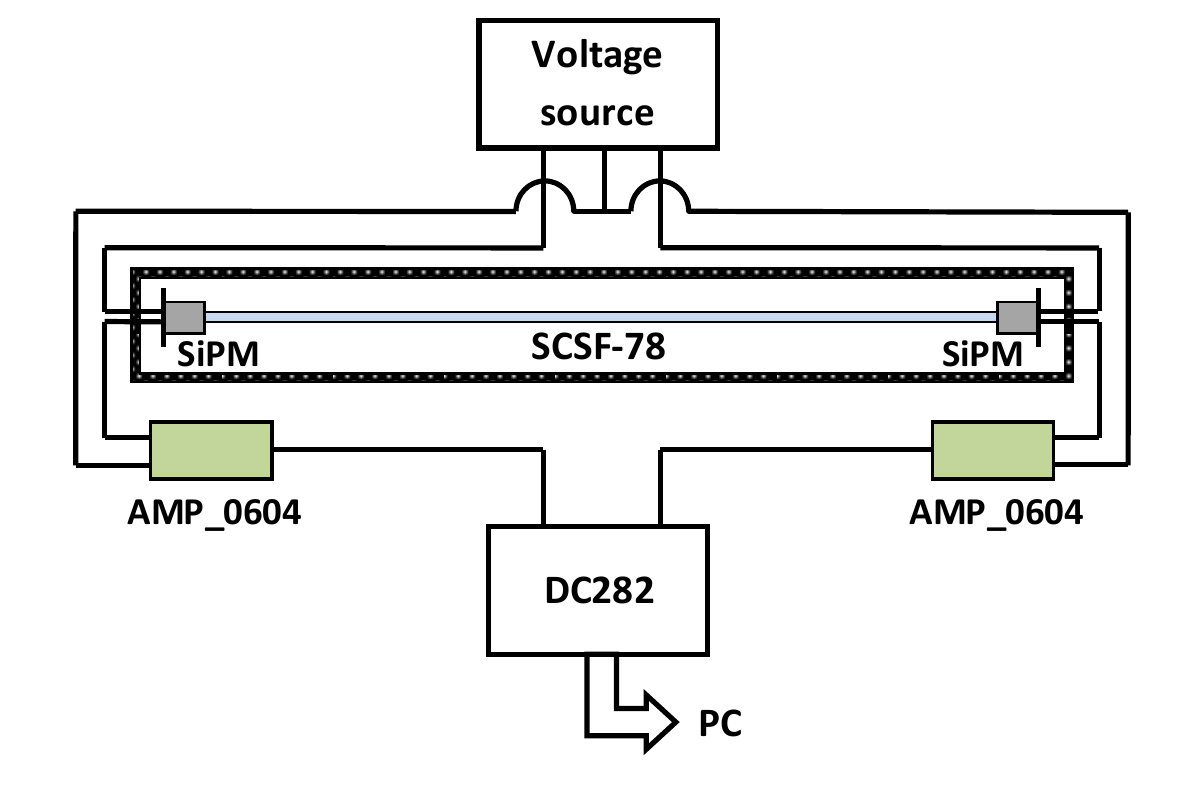}
\caption{Schematic block diagram of the experimental setup for measuring the coincidental timing distribution.}
\label{fig:fig11}
\end{figure}

For the initial verification of the proposed model an experimental setup from \autoref{fig:fig11} was employed, comprising a standard (Non-S) type SCSF-78 fiber from Kuraray \cite{fib10} with 1 mm$^2$ Photonique SSPM\_0611B1MM\_TO18 \cite{fib11} silicon photomultipliers as central components. The fiber itself is $\varnothing=0.83$ mm in outer diameter, with two layers of cladding and a spectral maximum of emitted light at 450 nm. The length of the selected fiber sample was $L$ = 0.52 m.  As a source of the scintillating pulses, a $\beta$-radioactive $^{137}$Cs sample was used. To ensure their protection from the environmental light, these elements were enclosed within the plastic container. The operating voltage for two diodes of different breakthrough values -- 26.4 V and 27.0 V -- was set to 27.6 V and 28.7 V, respectively. For a signal amplification the Photonique amplifiers AMP\_0604 \cite{fib11} were used.\\

\begin{figure}[t] 
\centering 
\includegraphics[angle=90,width=0.5\textwidth,keepaspectratio]{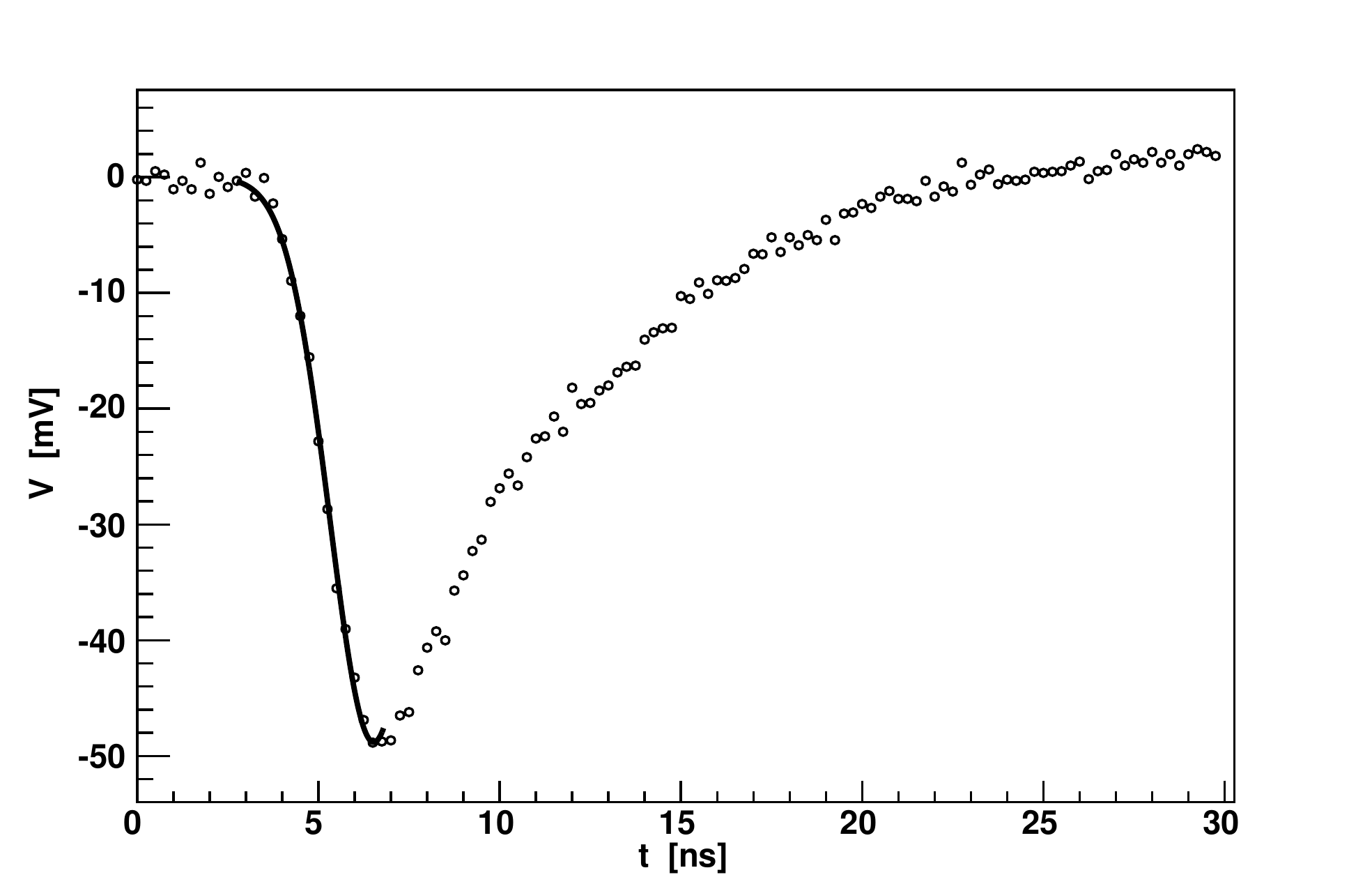}
\caption{An example of a signal recorded by DC282 digitizer, with discrete points sampled every 250 ps. A Gaussian fit of a leading edge is shown in full line.}
\label{fig:fig12}
\end{figure}

During the final stage of a signal transmission, amplified signals were fed to the fast 10-bit 4-channel Acqiris digitizer DC282 \cite{fib12}, with the maximal sampling rate of 8 GHz for a single active channel. With two active channels required for signal intake from both SiPM diodes, signals were recorded at 4 GHz sampling rate. Wholly stored for further offline analysis, signals may be freely and repeatedly accessed by any data extrapolation algorithm. At present time we are employing a simple baseline corrected calculation of an area below the voltage curve, i.e. signal integration, thus imitating the working principle of the charge collecting ADC units. For determination of the timing properties, specifically signal arriving and rising times, a constant fraction discrimination principle of classical TDC units was adopted. For this purpose, the leading edge of a signal is fitted to the Gaussian form from which the arrival time is extrapolated employing the constant fraction factor of $f_{CF}=20\%$. \autoref{fig:fig12} presents an example of the digitizer recording, together with a leading edge fit.\\

At all times during measurements a threshold equivalent to a trigger value of $\Delta_1=4$ for accepting events from one of the diodes was set above the range of a dark noise spectrum in order to avoid randomly triggered data collection. Subsequently, all the signals from the other diode were indiscriminately kept. Refined by a reduce in otherwise dominant dark noise due to the strict coincidence condition, an energy spectrum of a second diode was assumed to follow the multi-Poissonian form $\Phi(\chi)$ proposed in \cite{fib3}:
\begin{linenomath*}\begin{equation*}
\Phi(\chi)=\sum_{p=0}^{\infty}\sum_{s=0}^{\infty}\frac{e^{-\Pi}\Pi^p}{p!}\frac{e^{-p\Sigma}(p\Sigma)^s}{s!}\frac{\mathrm{exp}\left(-\frac{1}{2}\frac{(\chi-p-s)^2}{\sigma_0^2+(p+s)\sigma_1^2}\right)}{\sqrt{2\pi\left[\sigma_0^2+(p+s)\sigma_1^2 \right]}}
\tag{7.1}
\end{equation*}\end{linenomath*}
with $\chi=(x-x_0)/G$ as a number of fired pixels obtained by redefining the original energy scale $x$ in respect to a pedestal offset $x_0$ and pixel-to-pixel gain $G$. Summation indices $p$ and $s$ enumerate primary and secondary fired pixels\footnote{The firing of primary pixels is induced by an actual photon absorption, while that of secondary ones by a random dark noise occurrence, pixel-to-pixel crosstalk and after-pulsing \cite{fib13}. Therefore, with $\Pi_D$ as a mean number of the dark noise pulses within a signal integration interval, $\pi_C$ as a crosstalk probability and $\pi_A$ as an after-pulsing probability, the total mean $\langle p+s\rangle$ of fired pixels is given by:\begin{center}$\langle p+s\rangle=\Pi(1+\Sigma)=(\Pi+\Pi_D)(1+\pi_C)(1+\pi_A)$\end{center}} with $\Pi$ and $\Sigma$ as their mean, respectively. Parameters $\sigma_0$ and $\sigma_1$ define the pedestal and pixel noise. Fitting the data, values of \qquad\qquad $\Pi=(4.613\pm0.005)$ and $\Sigma=(0.467\pm0.002)$ were obtained.\\

For a comparison of the experimental and theoretical results, an emission probability distribution $E(t_E)$ from (4.1) was assumed, with the main component $\tau_1$ = 2.8 ns found in \cite{fib10}. The values of $r$ = 0.27 and $\tau_3$ = 14.2 ns were kept, while the last remaining parameter was set to $\tau_2$ = 1.1 ns. Though in \autoref{sec:chap4} a fiber model BCF-10 was selected due to its single cladding, validating the application of a model (4.4) for the path length dispersion $S_l(t_P)$, the same model was retained for SCSF-78 fiber -- a practice supported by an inner cladding thickness of only 3.4\% of a scintillating core diameter \cite{fib10}. Therefore, a light propagation speed inside the fiber is considered to be determined only by the core refractive index of $n_{core}=1.59$, while $n_{in}=1.49$ of inner cladding remains omitted within the simplified (4.3) model. However, a refractive index $n_{out}$ = 1.42 of outer cladding is essential for the definition of a maximal propagation time $t_{max}$ from (4.6). Finally, due to the rather short fiber sample used ($L$ = 0.52 m), a single exponential component of $\lambda$ = 4.27 m \cite{fib10} was adopted for an attenuation factor $A(t_P)$.\\

To obtain a measurement of the coincidental timing distribution, a non-collimated $^{137}$Cs source was placed at the center of the fiber: $l=L/2$. A single photon detection efficiency, addressed in \autoref{app:chapC}, was estimated to $\varepsilon_1 = \varepsilon_2$ = 20\% for both SiPM diodes \cite{fib11}. Based on the value $\Pi\approx4.6$ and the calculated photon survival probability $\Lambda_l$, a mean number of photons emitted during a scintillation decay was evaluated to $N=460$.\\

\begin{figure*}[t]
\centering
\begin{tabular}{cc}
\epsfig{file=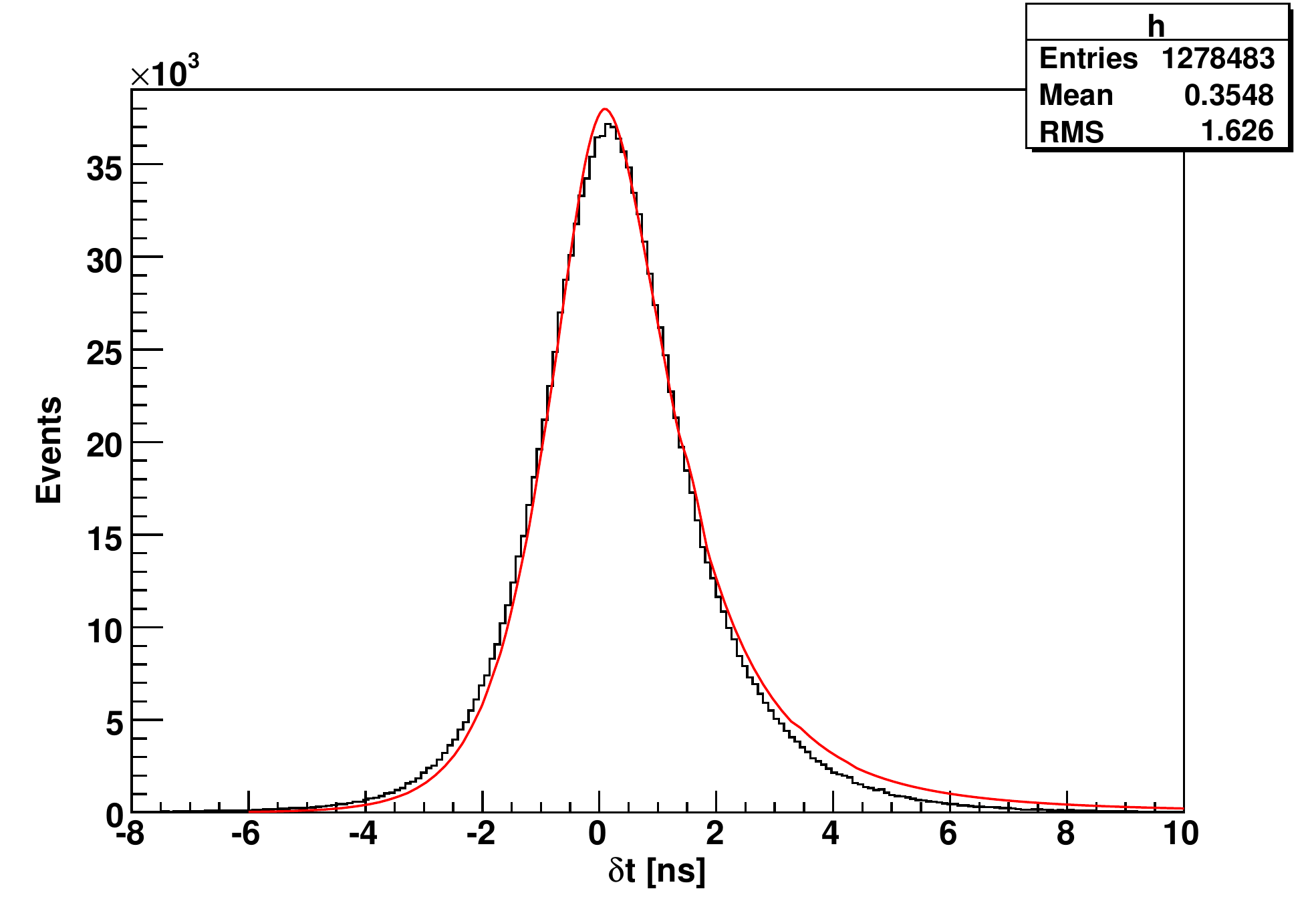,angle=90,width=0.5\textwidth,keepaspectratio,clip=} &
\epsfig{file=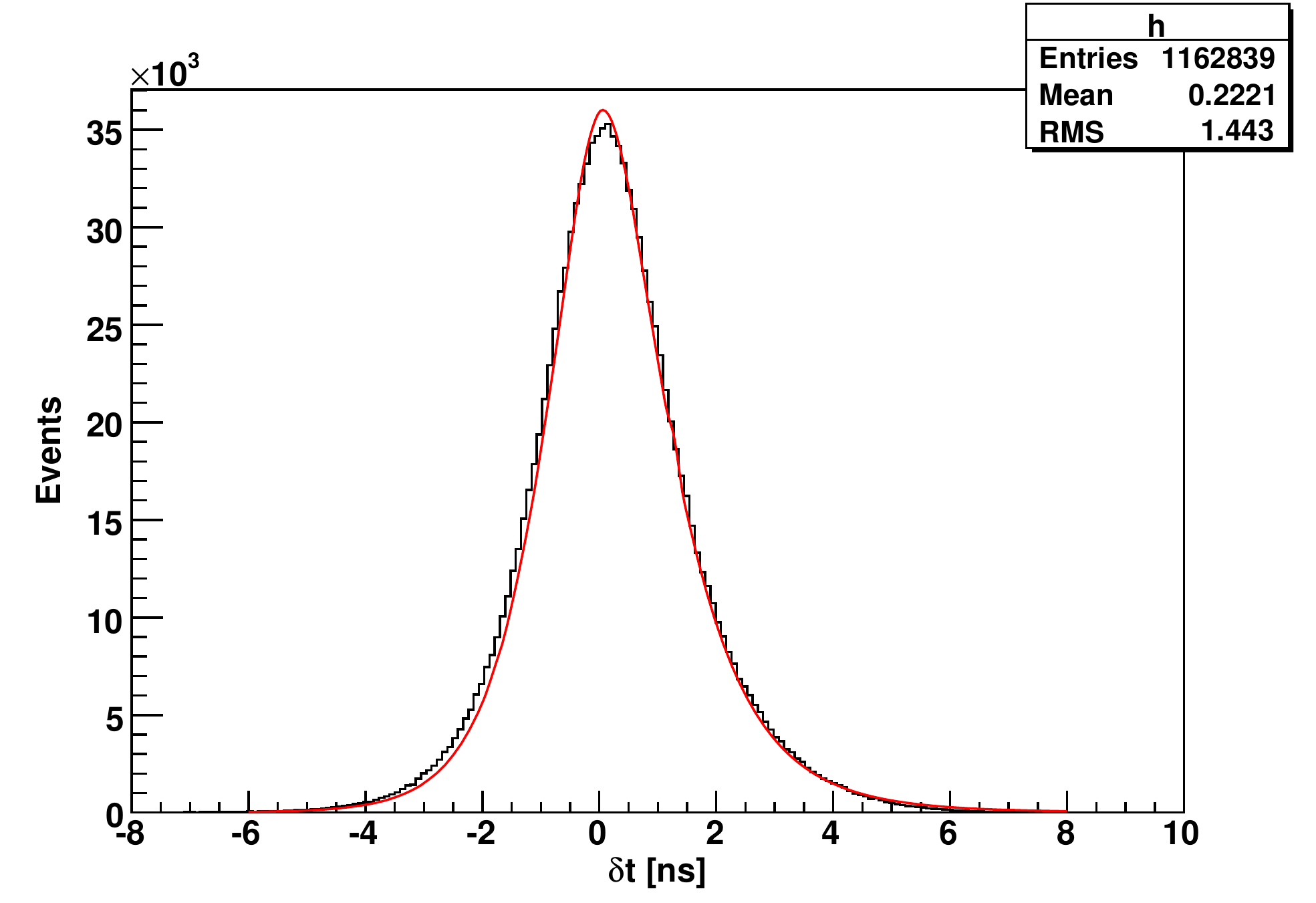,angle=90,width=0.5\textwidth,keepaspectratio,clip=} \\
$\Delta_2=0$ & $\Delta_2=2$ \\
\epsfig{file=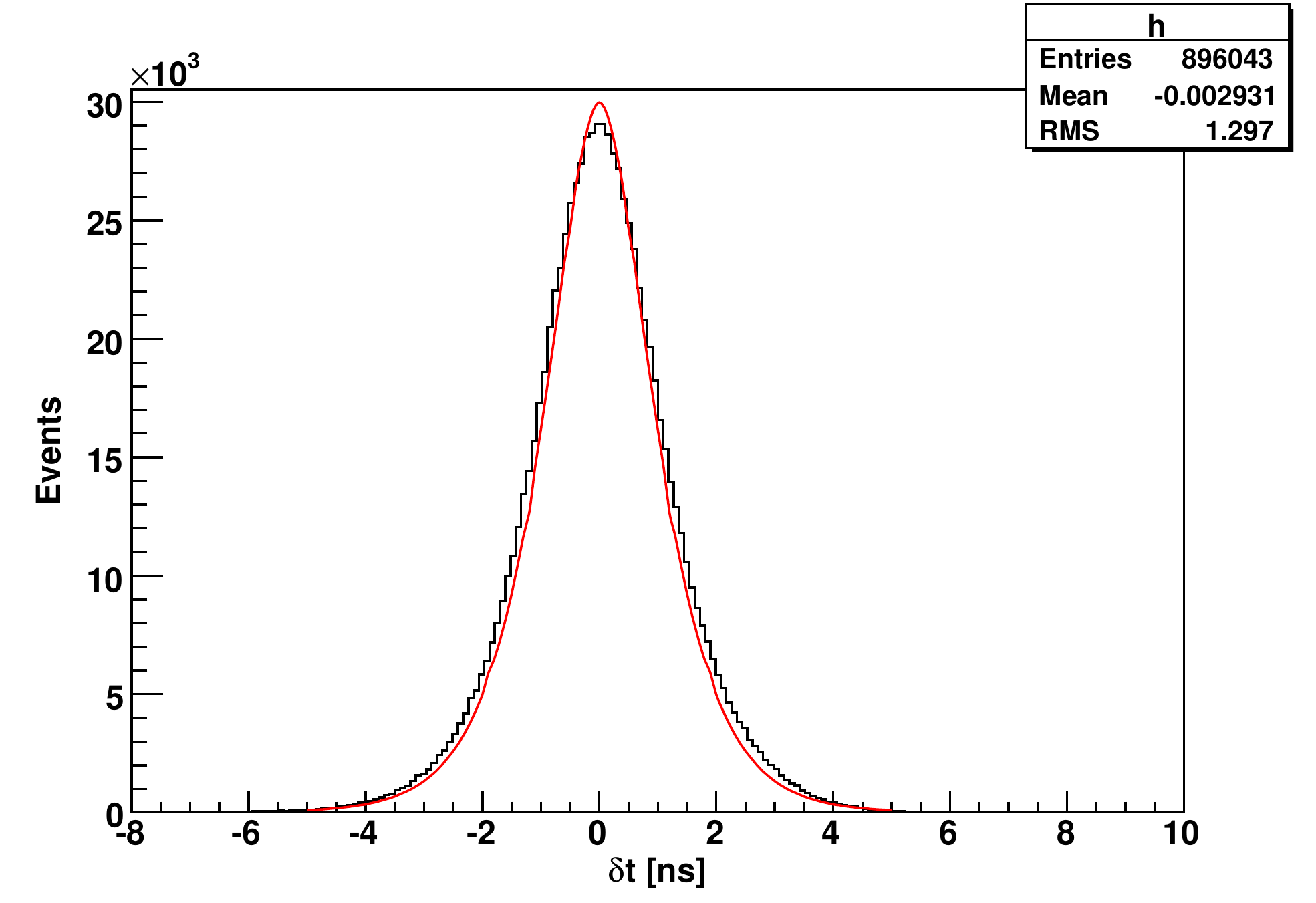,angle=90,width=0.5\textwidth,keepaspectratio,clip=} &
\epsfig{file=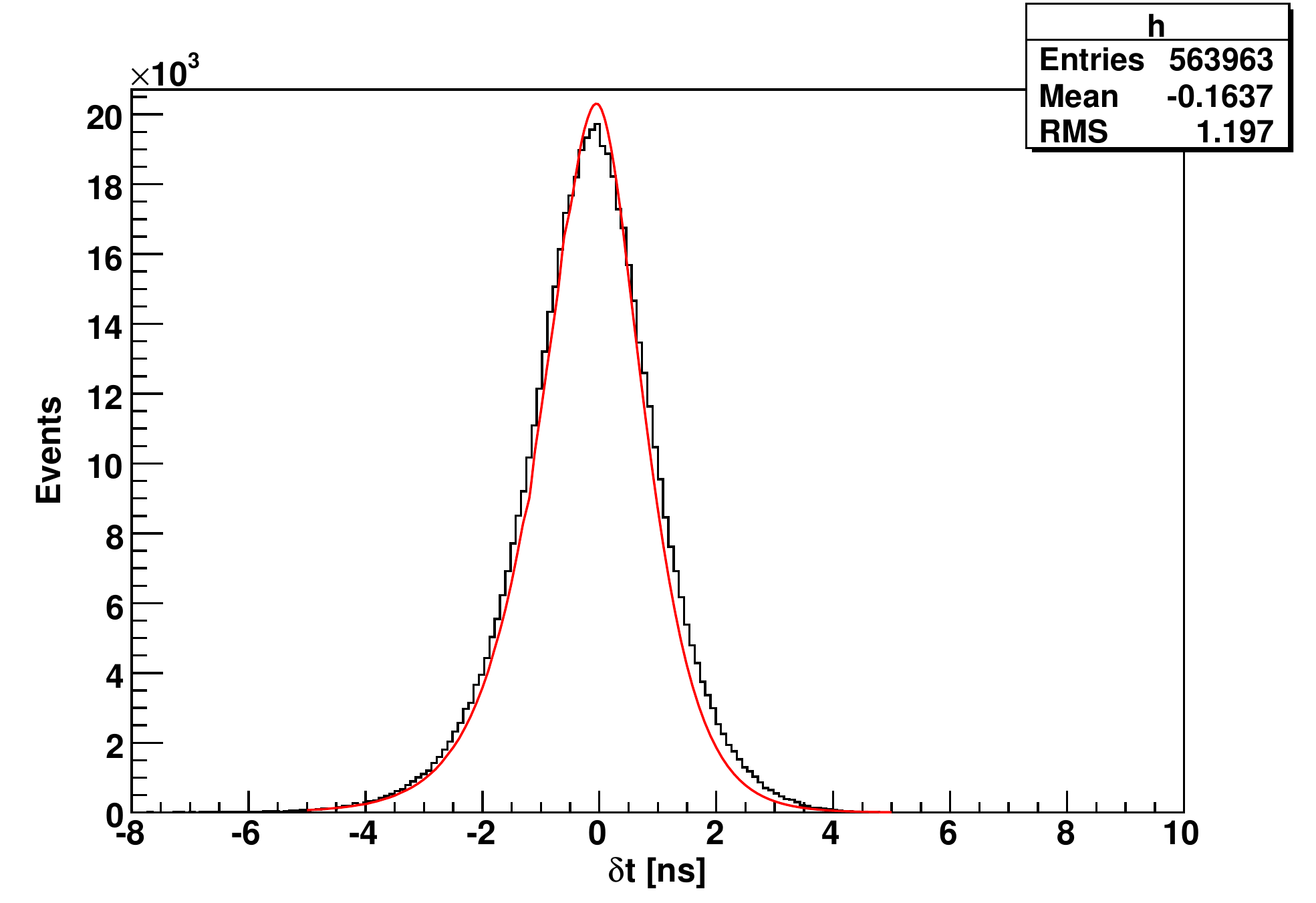,angle=90,width=0.5\textwidth,keepaspectratio,clip=} \\
$\Delta_2=4$ & $\Delta_2=6$ \\
\end{tabular}
\caption{Measured (histogram) versus calculated (smooth line) coincidental timing distributions for $\Delta_1=4$ and the different values of $\Delta_2$. FWHM values amount to 2.52 ns, 2.37 ns, 2.15 ns and 2.00 ns, respectively.}
\label{fig:fig13}
\end{figure*}

\autoref{fig:fig13} presents a coincidental timing distribution remained upon triggering with $\Delta_1$ = 4 and $\Delta_2$ = 0, 2, 4, 6. While a trigger value $\Delta_1$ was set in a course of the data collection, $\Delta_2$ was freely manipulated during offline analysis. Fair, if not excellent agreement between the experimental and theoretical results may be observed, regarding both the distribution width and a shape.\\

\begin{figure}[h!] 
\centering 
\includegraphics[width=0.5\textwidth,keepaspectratio]{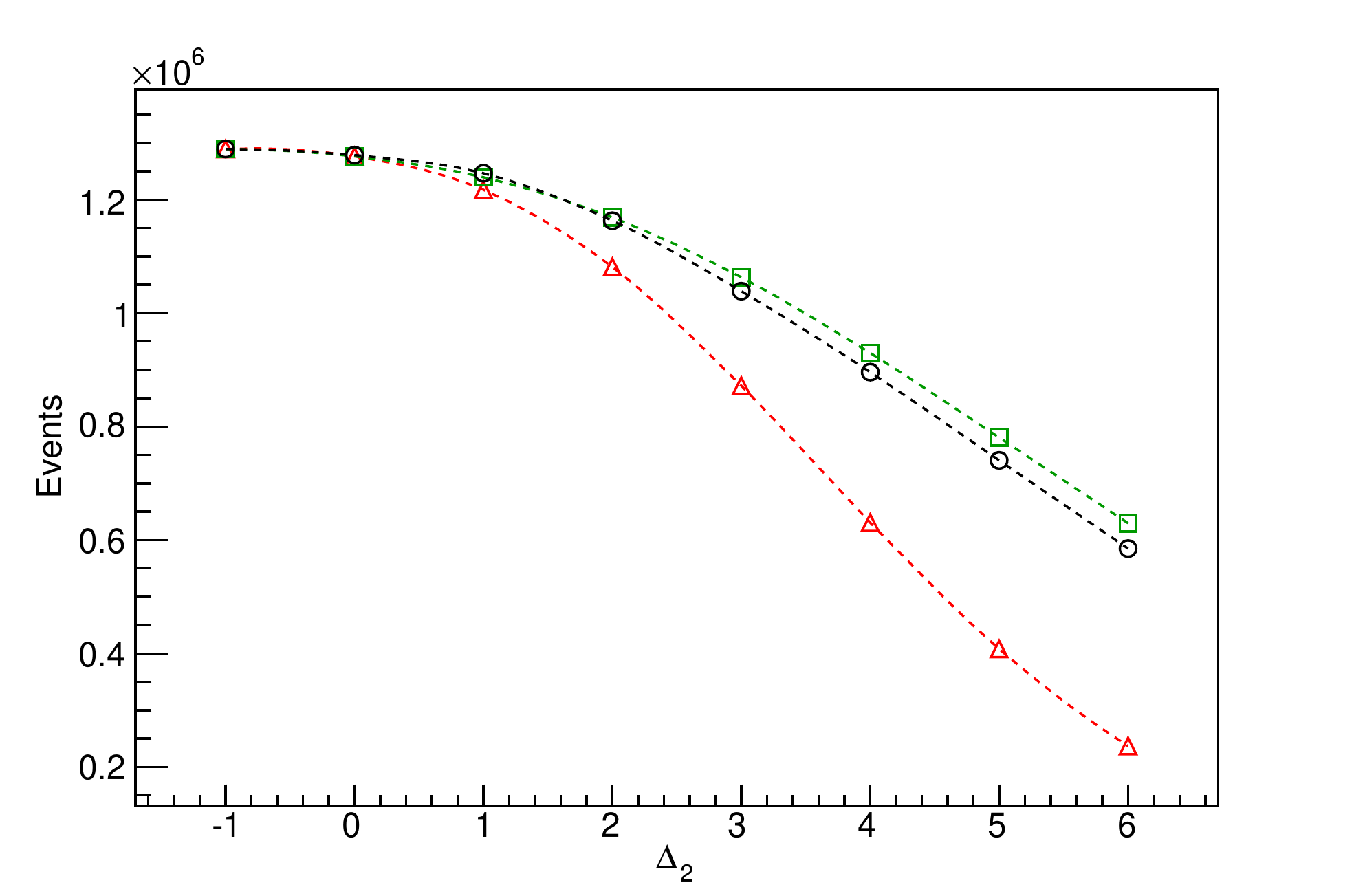}
\caption{Triggering induced decrease in a number of registered events, with a constant value of $\Delta_1$ = 4. Experimental points are shown as circles. Rectangle-shaped points are the result of an approximation (7.4), while the triangular ones of a model (7.2) devoid of the secondary pixel discharges. Dashed lines serve as a guide for an eye.}
\label{fig:fig14}
\end{figure}

The photon count considerations brought up in \autoref{sec:chap5} must be revised if they are to be translated into a pixel count, due to the presence of mechanisms inducing secondary pixel discharges, such as the dark noise, pixel-to-pixel crosstalk and after-pulsing \cite{fib13}. When considering only the primary discharges, i.e. those induced by an actual photon absorption, the probability $\psi_{\Pi}(\Delta)$ for more than $\Delta$ pixels firing may be expressed as \cite{fib4}:
\begin{linenomath*}\begin{equation*}
\psi_{\Pi}(\Delta)=1-\sum_{p=0}^{\Delta}\frac{e^{-\Pi}\Pi^p}{p!}
\tag{7.2}
\end{equation*}\end{linenomath*}
Therefore, a general trigger inclusive model (6.6) may be approximated as:
\begin{linenomath*}\begin{equation*}
\Psi_N(l;\Delta_1,\Delta_2)\approx\psi_{\Pi_l}(\Delta_1)\psi_{\Pi_{L-l}}(\Delta_2)
\tag{7.3}
\end{equation*}\end{linenomath*}
However, a secondary discharge occurrence somewhat modifies a purely Poissonian statistics, increasing the probability for pixel activation, extending (7.2) into:
\begin{linenomath*}\begin{equation*}
\psi_{\Pi\&\Sigma}(\Delta)=1-\sum_{p=0}^{\Delta}\sum_{s=0}^{\Delta-p}\frac{e^{-\Pi}\Pi^p}{p!}\frac{e^{-p\Sigma}(p\Sigma)^s}{s!}
\tag{7.4}
\end{equation*}\end{linenomath*}
\autoref{fig:fig14} plots the experimentally obtained decrease in a number of registered events against an approximate expression (7.4), together with the result of a secondary-discharge-free model (7.2).

\section{Conclusions}
\label{sec:chap8}
The scintillating fibers provide an efficient experimental tool for the detection of charged particles and at least a partial reconstruction of their trajectories. Due to the reconstructive methods relying upon the time of propagation of the scintillation light, timing properties of the scintillating fibers have become a central issue of their experimental performance. Therefore, understanding the sources of their clearly quantifiable coincidental timing resolution is a requirement of a fundamental importance for further experimental applications. From a developed model it was found that the rate of a scintillation decay, the number and spatial dispersion of emitted photons are defining factors for such resolution. Additionally, it was confirmed that resolution may indeed be improved by a triggering procedure, though implying the penalty of a reduced photon detection efficiency. Experimental results support the developed model, indicating, however, the necessity for a clear identification of a portion of random events within performed measurements. This is achieved by considering a physical impact of the relevant experimental elements besides the scintillating fiber itself, e.g. photon resolving devices. Further experimental investigation is planned in order to determine the practical applicability and limitations of such setup. \\\\\\*{\bf Acknowledgments}\\

This research was performed as a part of project 119-1191005-1021, with a leading researcher Dr. D. Bosnar, whose contribution to realizing this work is gratefully acknowledged.

Dr. P. Achenbach's supervision during the subject related IAEA fellowship CRO/10013 at Institut f\"{u}r Kernphysik, Johannes-Gutenberg Universit\"{a}t, Mainz, Germany is greatly appreciated. Furthermore, discussions with Dr. S. S\'{a}nchez Majos   are recognized as most useful.\\

%*{\bf Appendix}

\appendix
\section{Equivalency proof}
\label{app:chapA}
Given the general emission time probability distribution $f(t)$ -- previously denoted as $E(t_E)$ -- a probability distribution $\varepsilon(t_n)$ for the emission of the first of total $n$ photons at the moment $t_n$ is to be determined. There are two basic approaches to this procedure -- with or without arranging all the $n$ photons in time. A model including the time arrangement is defined by:
\begin{linenomath*}\begin{equation*}
\varepsilon(t_n)=n!f(t_n)\int\limits_{t_n}^\infty f(t_{n-1})\mathrm{d}t_{n-1}\int\limits_{t_{n-1}}^\infty f(t_{n-2})\mathrm{d}t_{n-2}\dotsb\int\limits_{t_2}^\infty f(t_{1})\mathrm{d}t_{1}
\tag{A.1}
\end{equation*}\end{linenomath*}
with the factorial term determining the number of photon permutations. Term $f(t_n)$ is a probability density for the emission of first photon at the moment $t_n$, while every subsequent integral regulates the probability for a given photon to be emitted following the previous one ($t_i>t_{i-1}$ for $(n+1-i)$-th photon). On the other hand, a model without arrangement requires:
\begin{linenomath*}\begin{equation*}
\varepsilon(t_n)=nf(t_n)\int\limits_{t_n}^\infty f(t_{n-1})\mathrm{d}t_{n-1}\int\limits_{t_{n}}^\infty f(t_{n-2})\mathrm{d}t_{n-2}\dotsb\int\limits_{t_n}^\infty f(t_{1})\mathrm{d}t_{1}
\tag{A.2}
\end{equation*}\end{linenomath*}
where initial $n$ regulates the number of possibilities to select only one (fist) photon, with all remaining $n-1$ following in any order. Hence, the difference between (A.1) and (A.2) is contained within combinatory factor and lower limits of integration. Equivalency between (A.1) and (A.2) would enable us to confidently and without reservations use whichever model proves to be more convenient for further calculations.\\

Thus, we begin the equivalency proof by considering the properties of a basic primitive function $F(t)$ for the emission probability distribution $f(t)$. The following holds:
\begin{linenomath*}\begin{equation*}
F(\infty)=0
\tag{A.3}
\end{equation*}\end{linenomath*}
This statement is supported by a physical requirement for the emission process to have a well defined mean lifetime\qquad\qquad\qquad $\langle t \rangle=\smallint_0^{\infty}tf(t)\mathrm{d}t$. For this to be true, it is obvious that $f(t)$ is required to asymptotically decrease strictly faster than $1/t^2$, directly implying that the primitive function $F(t)$ decreases faster than $1/t$, validating the claim (A.3). Moreover, from the probability normalization:
\begin{linenomath*}\begin{equation*}
1=\int\limits_0^\infty f(t)\mathrm{d}t=F(\infty)-F(0)
\tag{A.4}
\end{equation*}\end{linenomath*}
directly follows:
\begin{linenomath*}\begin{equation*}
F(0)=-1
\tag{A.5}
\end{equation*}\end{linenomath*}

Having obtained the result (A.3), it is easily shown that all the identical integrals in (A.2) give rise to the following solution for $\varepsilon(t_n)$:
\begin{linenomath*}\begin{equation*}
\varepsilon(t_n)=nf(t_n)\left[\int\limits_{t_n}^\infty f(t)\mathrm{d}t\right]^{n-1}=(-1)^{n-1}nf(t_n)F^{n-1}(t_n)
\tag{A.6}
\end{equation*}\end{linenomath*}
In order to prove that  a model (A.1) with photon arrangement yields the identical solution, a method of mathematical induction is applied. For this purpose we define the term $I_n(t_n)$ as:
\begin{linenomath*}\begin{equation*}
I_n(t_n)\equiv\int\limits_{t_n}^\infty f(t_{n-1})\mathrm{d}t_{n-1}\int\limits_{t_{n-1}}^\infty f(t_{n-2})\mathrm{d}t_{n-2}\dotsb\int\limits_{t_2}^\infty f(t_{1})\mathrm{d}t_{1}
\tag{A.7}
\end{equation*}\end{linenomath*}
and state that the following is valid:
\begin{linenomath*}\begin{equation*}
I_n(t_n)=\frac{(-1)^{n-1}}{(n-1)!}F^{n-1}(t_n)
\tag{A.8}
\end{equation*}\end{linenomath*}
Previous expression will serve as an induction hypothesis, supported by the basis:
\begin{linenomath*}\begin{equation*}
I_1(t_1)=1
\tag{A.9}
\end{equation*}\end{linenomath*}
Within the inductive step it is to be showed that the hypothesis holds for $I_{n+1}(t_{n+1})$:
\begin{linenomath*}\begin{equation*}
I_{n+1}(t_{n+1})=\int\limits_{t_{n+1}}^\infty f(t_n)I_n(t_n)\mathrm{d}t_n
\tag{A.10}
\end{equation*}\end{linenomath*}
Upon entering (A.8) into (A.10) and partially integrating:
\begin{linenomath*}\begin{align*}
\begin{split}
I_{n+1}&(t_{n+1})=\frac{(-1)^{n-1}}{(n-1)!}\int\limits_{t_{n+1}}^\infty f(t_n)F^{n-1}(t_n)\mathrm{d}t_n\\
&=\left[\begin{array}{ccc}
u=F^{n-1}(t_n) & \Rightarrow & \mathrm{d}u=(n-1)F^{n-2}(t_n)f(t_n)\mathrm{d}t_n \\
\mathrm{d}v=f(t_n)\mathrm{d}t_n &  \Rightarrow & v=F(t_n)\end{array}\right]\\
&\;\;\;\;\;\;\;\;\;\; =\frac{(-1)^{n-1}}{(n-1)!}\left[F^n(t_n)|_{t_{n+1}}^\infty-(n-1)\int\limits_{t_{n+1}}^\infty f(t_n)F^{n-1}(t_n)\mathrm{d}t_n\right]
\end{split}
\tag{A.11}
\end{align*}\end{linenomath*}
the appearance of the same integral may be noticed on both sides of the equation. Rearranging the terms leaves:
\begin{linenomath*}\begin{equation*}
I_{n+1}(t_{n+1})=\frac{(-1)^n}{n!}F^n(t_{n+1})
\tag{A.12}
\end{equation*}\end{linenomath*}
concluding the inductive proof.\\

Finally, the expression (A.1) is identified as a:
\begin{linenomath*}\begin{equation*}
\varepsilon(t_n)=n!f(t_n)I_n(t_n)
\tag{A.13}
\end{equation*}\end{linenomath*}
so that introducing (A.8):
\begin{linenomath*}\begin{equation*}
\varepsilon(t_n)=(-1)^{n-1}nf(t_n)F^{n-1}(t_n)
\tag{A.14}
\end{equation*}\end{linenomath*}
we may witness a result identical to (A.6).\\

Additionally, with only a step further it is shown that the probability distribution $\varepsilon(t_n)$ is, indeed, normalized, since:
\begin{linenomath*}\begin{equation*}
\int\limits_0^\infty \varepsilon(t_n)\mathrm{d}t_n=n!I_{n+1}(0)=(-1)^nF^n(0)=1
\tag{A.15}
\end{equation*}\end{linenomath*}
Here the identity (A.5) was used.

\section{Model justification}
\label{app:chapB}
Backtracking the origin of the expression (2.16) to (2.12), a particular explicit assumption of the proposed model becomes evident -- the numbers of photons (denoted by $n$) emitted towards the two ends of the fiber are considered independent, with a mean number $N$ of total emitted apparently being identified with a mean for every fiber end. Arguably, the more transparent approach would be to simultaneously consider both fiber ends during the emission of total $n$ photons, with $N$ as their genuine mean, while distributing them between those ends:
\begin{linenomath*}\begin{equation*}
\mathrm{P}_l=\sum_{n=2}^\infty \sum_{m=1}^{n-1} P_N(m)f_m(l)P_N(n-m)f_{n-m}(L-l)
\tag{B.1}
\end{equation*}\end{linenomath*}
It is to be noted that the probability density $\mathrm{P}_l$ for arrival of photons at both fiber ends requires a minimum of two photons to be emitted, while leaving at least one of them for every end. Upon entering (2.10) and (2.11) into (B.1), it may be readily shown that the following holds:
\begin{linenomath*}\begin{equation*}
\mathrm{P}_l=\rho_l\rho_{L-l}
\tag{B.2}
\end{equation*}\end{linenomath*}
which is the formal ground for the separate and independent treatment of fiber ends.

\section{Implementing advanced physical models}
\label{app:chapC}
Considerations contained within (4.3) and (4.6) are basically equivalent to the meridional approximation, which does not provide a complete description for the geometry of photon propagation \cite{fib7,fib9}. However, developed for a fully general selection of relevant distributions and parameters, a model comprised within (2.16) is readily available for incorporating any improvements and/or additions to the physical description from \autoref{sec:chap4}. Therefore, it is our goal to propose the manner in which to include a wide variety of advanced physical considerations. A more detailed overview of these contributions may be the subject of future work.\\

To improve upon the meridional approximation itself, a propagation time $t_P$ in (4.3) should be adequately redefined in terms of additional parameters, e.g. a skew angle, alongside primary $\theta$. Furthermore, such redefinition enables the inclusion of the multiple claddings via separation of the regimes for photons trapped inside a scintillating core from those passing through one or more claddings. Therefore, given a set $\{\theta^{(k)}\}$ of the relevant parameters, a propagation time should be defined as:
\begin{linenomath*}\begin{equation*}
t_P\left(\left\{\theta^{(k)}\right\}\right)=\sum_i t_i\left(\left\{\theta^{(k)}\right\}\right)\prod_k\Theta\left(\theta_i^{(k)}-\theta^{(k)} \right)\Theta\left(\theta^{(k)}-\theta_{i-1}^{(k)} \right)
\tag{C.1}
\end{equation*}\end{linenomath*}
with $\theta_i^{(k)}$ as a boundary of an $i$-th separate parameter space domain for the parameter $\theta^{(k)}$. Evidently, within (4.3)  it is assumed: $\theta^{(1)}\equiv\theta$, with $\theta_0^{(1)}=0$ and $\theta_1^{(1)}=\theta_{critical}$. In order not to omit a potential difference in the attenuation mode within separate fiber layers, an attenuation factor $A(t_P)$ should be redefined in parallel with $t_P$, in a manner identical to (C.1).\\

The following consideration brought up in \cite{fib7} is a Fresnel reflection above the critical angle for a total reflection. This effect is easily included by expanding the path length dispersion from (4.4) beyond previously imposed sharp cut, while setting $t_{max}\rightarrow\infty$. At this point it is to be noted that path length dispersion $S_l(t_P)$ inherits the normalization from its spatial origin, i.e. (4.2), making the integral $\smallint_0^{\infty} S_l(t_P)\mathrm{d}t_P$ not equal to unity, which is crucial for normalizing (2.16).\\

The finite transmittance, i.e. reflectance $q$ less than 1 even in a course of the otherwise total reflection, is to be implemented through the attenuation factor:
\begin{linenomath*}\begin{equation*}
A(t_P;q)=q^{\kappa(t_P)}A(t_P;q=1)
\tag{C.2}
\end{equation*}\end{linenomath*}
with $\kappa(t_P)$ as the number of reflections off the inner fiber interfaces \cite{fib7,fib9}. Reflections off the fiber ends present an additional matter that may be considered when describing the statistics of photon propagation. For this purpose an extended path length dispersion $S(t_P;l)$ may be constructed using the original reflections-free form $S_l(t_P)$:
\begin{linenomath*}\begin{equation*}
S(t_P;l)=(1-Q)\sum_{k=0}^{\infty}Q^{2k}\left[S_{l+2kL}(t_P)+QS_{2(k+1)L-l} (t_P)\right]
\tag{C.3}
\end{equation*}\end{linenomath*}
to enumerate the events with an even and odd number of reflections off the fiber ends prior to the final photon transmission. The reflectance $Q$ ($Q\neq q$) determines the probability for a given number of reflections as well as the final transmission probability ($1-Q$). Evidently, for $Q=0$ an extended dispersion $S(t_P;l)$ reduces to $S_l(t_P)$.\\

Many, if not all physical properties display a strong dependency on the emitted light wavelength $\lambda$ -- from the photon emission itself, to refractive indices of the fiber, attenuation lengths, etc. Given the normalized wavelength emission spectrum $p(\lambda)$, this dependency is to be included by averaging the probability density $\rho(t_P,t_E)$ from within (2.15):
\begin{linenomath*}\begin{equation*}
\rho(t_P,t_E)=\int\limits_{\lambda_{min}}^{\lambda_{max}}\rho(t_P,t_E;\lambda)p(\lambda)\mathrm{d}\lambda
\tag{C.4}
\end{equation*}\end{linenomath*}
over the relevant wavelength interval extending from $\lambda_{min}$ to $\lambda_{max}$.\\

During the charged particle's transition through the fiber material, an entire array of scintillating molecules is excited along its path, which generally may not be assigned a unique value for the distance $l$ from one of the fiber ends. Thought the basic principle for including photon origin variations remains the same as for wavelength variations, it requires somewhat more complex approach. Therefore, let us assume that for every single particle passing through the fiber, their trajectory is parameterized by $X$ -- a single parameter or, symbolically, a set of parameters. For a collimated particle beam this parameterization may be omitted. However, its significance becomes apparent, for example, in a case of a non-collimated isotropically emitting particle source. When including this improvement in respect to the point-like description of scintillating pulses, the first step is to translate a probability distribution $D_l(t_A)$ from (2.15) into a form $\mathpzc{D}(t_A;X)$:
\begin{linenomath*}\begin{equation*}
\mathpzc{D}(t_A;X)=\int\limits_{l_{min}(X)}^{l_{max}(X)}D_{l(X)}\left(t_A-t_{l(X)}\right)\mathrm{d}l(X)
\tag{C.5}
\end{equation*}\end{linenomath*}
taking into account that $l$ varies from $l_{min}(X)$ to $l_{max}(X)$. There is an additional subtle consideration to be noted -- the inclusion of the moment $t_{l(X)}$ of a particle reaching the distance $l(X)$ from one of the fiber ends. Its definition may be as simple as:\qquad $t_{l(X)}=[l(X)-l_0(X)]/v_l$ with $l_0(X)$ as a value for $l(X)$ at the moment of particle entering the fiber and $v_l$ as its speed along the fiber axis. The following step:
\begin{linenomath*}\begin{equation*}
\mathpzc{R}(\delta t;X)=\int\limits_0^{\infty}\mathpzc{D}(t_A;X)\mathpzc{D}(t_A+\delta t;X)\mathrm{d}t_A
\tag{C.6}
\end{equation*}\end{linenomath*}
brings nothing new to the practice from (2.16), save for inheriting the $X$-dependency from $\mathpzc{D}(t_A;X)$. However, it constitutes only an intermediate step preceeding the final calculation of a coincidental timing distribution $R_{\langle l\rangle}(\delta t)$:
\begin{linenomath*}\begin{equation*}
R_{\langle l\rangle}(\delta t)=\int\limits_{X_{min}}^{X_{max}}\mathpzc{R}(\delta t;X)p(X)\mathrm{d}X
\tag{C.7}
\end{equation*}\end{linenomath*}
which is -- upon averaging procedure considering a distribution $p(X)$ -- parameterized by a mean value $\langle l\rangle$. At this point an important note has to be made regarding an integration being applied to $D_l(t_A)$ in (C.5), while in (C.7) to $\mathpzc{R}(\delta t;X)$. The practice from (C.5) allows the photons reaching separate fiber ends to be assigned different origins $l(X)$, with a practice from (C.7) compelling all the photons from a single scintillation pulse to be assigned a single particle trajectory.\\

In the end, even the properties of the light detection devices may be implemented through a proposed model. A single photon detection efficiency $\varepsilon_i$ for a detector at $i$-th fiber end is to be included in a simple manner:
\begin{linenomath*}\begin{equation*}
A(t_P;\varepsilon_i)=\varepsilon_iA(t_P;\varepsilon_i=1)
\tag{C.8}
\end{equation*}\end{linenomath*}
Furthermore, in a course of the signal timing measurements a spread $\sigma(t_A)$ due to any kind of fluctuations within a light detecting system -- electronic noise, fluctuations during the electric charge collection, variations in a rise time of a signal, etc. -- may be taken into account. Consequently, it is manifested through the modified arrival times distribution $\mathscr{D}_l(t_A)$:
\begin{linenomath*}\begin{equation*}
\mathscr{D}_l(t_A)=(D_l*\sigma)(t_A)=\int\limits_{-\infty}^{\infty}D_l(t)\sigma(t_A-t)\mathrm{d}t
\tag{C.9}
\end{equation*}\end{linenomath*}
gained by means of a functional convolution.

\end{document}